\documentclass[sigconf]{acmart}
\usepackage{subcaption}
\usepackage{cleveref}
\usepackage{tabularx, colortbl}
\usepackage{booktabs}
\usepackage{contour}
\usepackage{soul}
\usepackage{fontawesome5}
\usepackage{xspace}
\usepackage{enumitem}
\usepackage{transparent}
\usepackage{multirow}

\contourlength{0.8pt}
\setuldepth{hello}

\newcounter{theory}

\newcolumntype{L}[1]{>{\raggedright\let\newline\\\arraybackslash\hspace{0pt}}m{#1}}

\newcommand{\myuline}[1]{%
  \ul{{\phantom{#1}}}%
  \llap{\contour{white}{#1}}%
}

\crefname{figure}{Figure}{Figures}

\newcommand{\newText}[2]{#1}

\definecolor{pastelgray}{HTML}{efefef
}
\definecolor{headergray}{HTML}{efefef
}
\definecolor{resources}{HTML}{4477AA}
\definecolor{training}{HTML}{EE6677}
\definecolor{impact}{HTML}{CCBB44}
\definecolor{personalharms}{HTML}{228833}
\definecolor{minus}{HTML}{C00000}
\definecolor{plus}{HTML}{4F6228}

\AtBeginDocument{%
  \providecommand\BibTeX{{%
    \normalfont B\kern-0.5em{\scshape i\kern-0.25em b}\kern-0.8em\TeX}}}

\copyrightyear{2024}
\acmYear{2024}
\setcopyright{cc}
\setcctype[4.0]{by}
\acmConference[CHI '24]{Proceedings of the CHI Conference on Human Factors in Computing Systems}{May 11--16, 2024}{Honolulu, HI, USA}
\acmBooktitle{Proceedings of the CHI Conference on Human Factors in Computing Systems (CHI '24), May 11--16, 2024, Honolulu, HI, USA}
\acmDOI{10.1145/3613904.3642025}
\acmISBN{979-8-4007-0330-0/24/05}

\begin{document}

\title{
Counterspeakers' Perspectives: Unveiling Barriers and AI Needs in the Fight against Online Hate
}


\author{Jimin Mun}
\email{jmun@andrew.cmu.edu}
\affiliation{%
  \institution{Language Technologies Institute, Carnegie Mellon University}
  \city{Pittsburgh}
  \state{PA}
  \country{USA}
}

\author{Cathy Buerger}
\authornote{Equal contribution.}
\affiliation{%
  \institution{Dangerous Speech Project}
  \city{Washington, D.C.}
  \country{USA}
}
\author{Jenny T. Liang}
\authornotemark[1]
\affiliation{%
  \institution{Human-Computer Interaction Institute Carnegie Mellon University}
  \city{Pittsburgh}
  \state{PA}
  \country{USA}
}
\author{Joshua Garland}
\affiliation{%
  \institution{Arizona State University}
  \city{Tempe}
  \state{AZ}
  \country{USA}
}
\author{Maarten Sap}
\affiliation{%
  \institution{Language Technologies Institute, Carnegie Mellon University}
  \city{Pittsburgh}
  \state{PA}
  \country{USA}
}

\renewcommand{\shortauthors}{Mun, et al.}
\renewcommand{\shorttitle}{Counterspeakers’ Perspectives: Unveiling Barriers and AI Needs}

\begin{abstract}
    Counterspeech, i.e., direct responses against hate speech, has become an important tool to address  the increasing amount of hate online while avoiding censorship. Although AI has been proposed to help scale up counterspeech efforts, this raises questions of how exactly AI could assist in this process, since counterspeech is a deeply empathetic and agentic process for those involved. 
    In this work, we aim to answer this question, by conducting in-depth interviews with 10 \newText{extensively}{R3} experienced counterspeakers and a large scale public survey with 342 \newText{everyday social media users}{R3}. 
    In participant responses, we identified four main types of barriers and AI needs related to resources, training, impact, and personal harms. However, our results also revealed overarching concerns of authenticity, agency, and functionality in using AI tools for counterspeech. To conclude, we discuss considerations for designing AI assistants that lower counterspeaking barriers without jeopardizing its meaning and purpose.
\end{abstract}

\ccsdesc[500]{Human-centered computing~Empirical studies in HCI}


\keywords{counterspeech, hate speech, AI-supported counterspeech, AI-mediated communication, online activism}




\maketitle

\section{Introduction}

\textit{Counterspeech}, i.e., direct responses to mitigate the harms of hateful or dangerous speech \cite{beneschConsiderationsSuccessfulCounterspeech2016}, has emerged as a more positive, community-oriented alternative to deletion-based content moderation that avoids censorship concerns \cite{myers2018censored}.
Its goals are multifaceted; counterspeech aims to not only minimize harms of hateful speech, but also to promote positive changes in online communities through open dialogue among users \cite{buergerWhyTheyIt2022a} and by fostering a sense of community \cite{buergerIamhereCollectiveCounterspeech2021}.
Furthermore, how counterspeech is done can be highly varied and complex, as it
can range from individual replies to a hateful post to coordinated mass responses via organized groups and hashtags (e.g., \#iamhere, \#BlackLivesMatter, \#stopasianhate) \cite{buergerIamhereCollectiveCounterspeech2021,yangNarrativeAgencyHashtag2016,bing2021asian,friessCollectiveCivicModeration2021,garland-etal-2020-countering}.
As hate speech prevalence grows in online spaces \cite{Leetaru2019-xf,Baggs2021-cl,sahaRiseFearSpeech2023}, coordinating and responding with counterspeech has become increasingly challenging \cite{CHUNG2021100150,citron2011intermediaries}.



AI has recently emerged as a potential tool or solution \cite{cypris2022intervening,ashida-komachi-2022-towards,Zhu2021-prune} to assist with this increased demand for counterspeech.
However, designing an AI system for counterspeech is a unique challenge that requires an understanding of larger context of its impact on those who do it \cite{blytheAntiSolutionistStrategiesSeriously2016}.
On one hand, it is important to reduce the burden on counterspeakers \cite{Howard2021-ll}, who are sometimes victims of hate speech themselves \cite{cepollaro_counterspeech_2023}.
On the other hand, naive AI solutions that simply generate counterspeech without human oversight could significantly detract from the authentic, empowering, and emotionally connecting experience that counterspeech provides to users \cite{friessCollectiveCivicModeration2021,buergerIamhereCollectiveCounterspeech2021}, despite being a growing area of research in AI \cite{chung2023understanding,Zhu2021-prune,Saha2022-countergedi}.

\newText{Given these challenges and tensions, 
a more deliberate approach that involves stakeholders with varying degrees of counterspeaking experience is required, to design tools that can empower counterspeakers and increase participation in counterspeech.
To enable this, we conduct the first set of studies to}{R1}
\newText{ 
collect various perspectives from users}{2AC} to understand their counterspeech experiences and to inform the possible role of AI technology in counterspeech.
We partnered with an NGO specialized in responses to hate speech to 
ask the following research questions:
\begin{enumerate}
    \item [\textbf{RQ1}] What are the barriers, if any, for users to engage in counterspeech?
    \item [\textbf{RQ2}] What AI tools, if any, can assist in removing or lowering these barriers?
    \item [\textbf{RQ3}] What are user concerns of using AI tools for counterspeech?
\end{enumerate}
We conduct our studies with two populations of participants with differing levels of counterspeech experience to understand both the needs of current, \newText{extensively experienced counterspeakers as well as less-experienced everyday social media users}{R3}.
We conducted semi-structured long-form interviews with activists who regularly respond to hatred online (\textit{N=10}), 
and also surveyed with \textit{342} everyday social media users who may have never participated in counterspeech. 



Through qualitative analysis, such as grounded theory, we found several emerging themes, corroborated by our quantitative results.
Our participant responses surfaced tension between barriers and motivations to counterspeech. 
At a high level, participants identified \textit{limited resources}, \textit{lack of training}, \textit{unclear impact}, and \textit{fear of personal harms} as deterrence to engage, but were motivated by a sense of \textit{moral duty} and \textit{positive impact}.
We characterized AI tools that were envisioned by participants, functional requirements that connected to the barriers and the characteristics of such tools (e.g., emotional, factual, empowering), toward designing participatory AI systems for counterspeech. 
Furthermore, we found that while participants thought that AI tools could help address these barriers, they expressed \textit{functional doubts} and strong concerns of \textit{authenticity} and \textit{agency} in using AI counterspeech tools.



These findings highlight the gap between current research directions in counterspeech AI assistance and designing an empowering tool for counterspeakers in addressing the concerns raised by our participants.
As a step towards closing this gap, we outline design considerations for \textit{authenticity of counterspeech}, \textit{moral agency}, and \textit{mental health} by connecting our findings to previous works. 
More specifically, we make recommendations towards ensuring transparency in online communication, avoiding moral passivity and disengagement, and promoting mindfulness and intentionality in interactions and call for future studies to explore the concerns of AI counterspeech assistance and their solutions.

\newText{
In summary, our work seeks to understand from both experienced activist counterspeakers and lay-users their perspectives on online counterspeech through learning about their counterspeech experiences, envisioned AI tools, and concerns about such tools.
In so doing, we contribute to the discussion on the fight against online hate especially on counterspeech and role of AI-driven counterspeech tools.
Specifically, our contributions are in \textbf{(1)} synthesizing the experience of two different populations of users to describe a theory of counterspeech engagement and barriers to counterspeech, \textbf{(2)} collaboratively envisioning potential usage of AI, \textbf{(3)} surfacing concerns of AI tools, and \textbf{(4)} broadening the vision of AI tools for counterspeech and offering design considerations.}{2AC,R1}
\section{Related Work}
To explore counterspeaker needs and AI tools for counterspeech, we first situate counterspeech more broadly in relation to other hate speech responses (Section~\ref{ssec:rw-hatespeech}).
We then investigate related works to understand counterspeech and counterspeakers (Section~\ref{ssec:rw-counterspeech}).
Finally, we provide an overview of prior works in AI and counterspeech (Section~\ref{ssec:rw-ai}) to investigate the direction of AI assistance in counterspeech and the research gaps that we aim to address in our work.


\subsection{Hate Speech Responses}
\label{ssec:rw-hatespeech}
Hate speech is a commonly seen problem in online communication, especially on social media platforms \cite{mondal2017measurehate,das2020hate}. 
\newText{
Definition of hate speech varies across countries, institutions, and communities, which leads to disagreement over appropriate responses \cite{kulenovicShouldDemocraciesBan2022}.
In the most extreme cases, hate speech is defined as a weaponized speech that causes direct harms to individuals such as inciting violence or hostility \cite{kulenovicShouldDemocraciesBan2022}; however, speech that causes indirect harms including devaluation of community norms such as inclusivity can also be considered hate speech \cite{kulenovicShouldDemocraciesBan2022,parekhThereCaseBanning2012}. 
Thus the many forms and degrees of harm in hate speech require flexible countering responses \cite{jurgens-etal-2019-just}. 
One aspect of hate speech most commonly shared by various definitions is that it has a specific target, a group of people distinguished by different identities such as religion, race, gender, and sexual identity \cite{fortunaSurveyAutomaticDetection2018}.}{R3}
With the scale and pervasiveness of cyberspaces, online hate speech can be especially damaging to the targeted social groups as it fosters hostility and perpetuates stereotypes and can even lead to off-line violence \cite{castano-pulgarinInternetSocialMedia2021,buergerSpeechDriverIntergroup2021}. 

One of the most prevalent solutions to the growing scale of hate speech has been content moderation through algorithmic censorship \cite{Poletto2021,fortunaSurveyAutomaticDetection2018}.
However, algorithmic censorship has been shown to have many shortcomings \cite{parker2023detection}, causing growing concerns over freedom of speech, especially with the algorithms tailored to serve the platform often excluding the users from its decisions \cite{kaye2019speechpolice,cobbeAlgorithmicCensorshipSocial2021,areShadowbanCycleAutoethnography2022}.
Moreover, many automated hate speech detection has been shown to be inaccurate, unable to detect subtle hate speech \cite{han-tsvetkov-2020-fortifying,hartvigsen2022toxigen} or biased against certain communities causing further marginalization \cite{sap2019risk,sap-etal-2022-annotators}.
Some new proposed methods of online governance include user-based and community-based moderation models \cite{seering2020moderation} such as collective-decision moderation \cite{2022-perceived-legitimacy}, which require more involvement from users but offer a more democratic way to make moderation decisions.
Counterspeech, along with these moderation methods, provides users and platforms with a more flexible solution that can be used as an alternative or complement to deletion-based methods to counter hate while influencing a positive cultural shift through dialogue.


\subsection{Counterspeech and Counterspeakers}
\label{ssec:rw-counterspeech}
Counterspeech is a complex phenomenon with many potential benefits to correct stereotypes \cite{lemeire2021falsify}, prevent the spread of misinformation \cite{langton_blocking_2018}, reinforce correct information \cite{siebert2023mitigating}, and to expand responses to even more covert forms of hate speech \cite{baiderAccountabilityIssuesOnline2023}.
Its beneficial outcomes can help alleviate online hate in many contexts: as part of a moderation decision to increase transparency by communicating moderation intent \cite{sasse2023moderationtype}, as an intervention against cyberbullying that also shows support towards the victims \cite{rudnickiSystematicReviewDeterminants2023}, and more broadly, as a remedy for harmful speech (e.g., misinformation, propaganda, hateful speech) \cite{nunziatoVarietiesCounterspeechCensorship2020}.  
Research on online counterspeech has been focused on social media interactions \cite{miskolciCounteringHateSpeech2020,nunziatoVarietiesCounterspeechCensorship2020} to measure its patterns, effectiveness, and role against hate speech.
Many of these studies have been focused on finding effective content for belief or sentiment change measured through subsequent behaviors of the poster (e.g., deletion of the hateful post) or those engaged in the discourse (e.g., sentiment of the comments) and have shown empathy \cite{hangartner_empathy_2021} and civility \cite{hanCivilityContagiousExamining2018} to have some positive impact on these measures.
However, it requires effort and engagement from the users doing it, which becomes challenging.

Counterspeakers' goals are often multifaceted: to change the view of the bystanders, to recruit more counterspeakers, or to strengthen community norms \cite{buergerWhyTheyIt2022a}.
Moreover, counterspeaking can create connection and solidarity against hate among users as they respond to online hate speech collectively as a part of an organized movements such as \#iamhere \cite{buergerIamhereCollectiveCounterspeech2021,friessCollectiveCivicModeration2021}. 
\newText{While content moderators sometimes use counterspeech to de-escalate conversations and algorithmic support for such proactive moderations have been investigated (e.g., flagging conversations that are likely to devolve into toxicity \cite{schluger2022proactive}), most counterspeakers who are everyday users have different positionality and power compared to the moderators \cite{Weld_Zhang_Althoff_2022}.
Therefore, in our work, we focus on counterspeakers and the public to understand the barriers to counterspeaking and tools to support counterspeech.
More specifically, we look at hurdles and reasoning behind decisions to not engage and how AI tools can lower these barriers, which have not yet been studied and require further investigation.}{R1}

\subsection{AI Assistance in Counterspeech}
\label{ssec:rw-ai}
As counterspeech has become an increasingly important avenue for responding to hate speech online, the issue of scaling this effort has become an active research topic. 
\newText{Full automation of counterspeech is thought of as artificial agents such as bots responding to speech that are detected to be hateful \cite{cypris2022intervening}. 
Following such frameworks, automation of hate speech detection could be considered a part of counterspeech assistance; however, the focus of these studies have been on detection only with limited annotations \cite{fortunaSurveyAutomaticDetection2018,Poletto2021} lacking in specification or adaptation for diverse solutions \cite{parker2023detection}.}{}
Counterspeech generation and detection have more recently become the focus of research community. 
Automatic detection \cite{yuHateSpeechCounter2022} and computational analysis of large scale counterspeech \cite{garland-etal-2020-countering} have been used to understand its characteristics and to inform effective content.
Prior works on automatic generation of counterspeech \cite{Zhu2021-prune,Mathew2019-thou} relied on curated or scraped datasets \cite{chung-etal-2019-conan,hassan2023discgen} and evaluation metrics based on correct countering claims \cite{halimWokeGPTImprovingCounterspeech2023} or emotion and politeness \cite{Saha2022-countergedi}.
Some methods used limited response intent such as question, denouncing, and humor in dataset \cite{chung-etal-2019-conan} and as part of the generation method \cite{gupta2023counterspeeches}.
\newText{In addition, counterspeech generation in dialogue systems (e.g., Alexa, Cortana, etc.) has come into focus as well \cite{ullmann2023counterspeech}.
These approaches, while offering interesting explorations, did not consider counterspeech and its full context or the complex user intentions of counterspeaking.}{}

\newText{Moreover, ethical implication and value of automation have not been addressed and require further consideration \cite{gillespieContentModerationAI2020,cypris2022intervening}.
This is becoming an increasingly important topic to address as online hate incidents are increasing \cite{OnlineHateHarassment} and AI, which is not well defined even among experts \cite{2016ReportOne}, is becoming more and more prevalent in our society and in the public's daily activities \cite{saksPublicAwarenessArtificial2023}. }{}
Therefore, in this work, we seek to answer how AI
can best help counterspeakers by bringing attention to users, their intentions and barriers, and collaboratively identifying possibilities and concerns of AI for counterspeech. 


\section{Methodology}

To understand counterspeech with diverse perspectives of both activists and everyday users, we used a mixed methods approach \cite{Schoonenboom2017}.
\newText{
The interview participants, who are counterspeech activists, are characterized by their extensive involvement in counterspeaking through systematic efforts (e.g., repetitive, over an extended period of time) and their understanding of counterspeech.
These participants were recruited through pre-established relationships with the partnered NGO. 
On the other hand, the survey participants were recruited on Amazon Mechanical Turk (MTurk) without particular pre-screening based on their counterspeech experiences, to capture perspectives more representative of everday social media users. 
}{R3}
Our approach focuses on understanding both the depth and breadth of counterpseech experiences by mixing qualitative and quantitative studies of the two populations.
To this end, we developed semi-structured interview guidelines and a survey informed by the responses using an exploratory sequential design \cite{creswell2017designing}.
\newText{Moreover, to gather diverse applications of AI from both groups of participants, we first ask questions about an envisioned application without defining the type of AI (e.g., generative AI, classification models) as to ensure creative and even future-oriented answers. 
We then ask more specific questions about some common types of AI tools envisioned by experts in relevant literature and by the research team, such as tools that use generative AI (e.g., counterspeech bots, revision support through rewriting) or uses retrieval based support (e.g., tool that suggests facts).}{2AC \& R3}
We used grounded theory to analyze the qualitative results and triangulation to support our findings discussed in Section~\ref{sec:findings}.

\begin{table}[!ht]
    \centering
    \footnotesize
    \begin{tabular}{c|cc}
        \toprule
        Participant ID & Country of Residence & Years Counterspeaking \\
        \hline
        1 & Ethiopia & 2\\
        2 & USA & 4\\
        3 & Cameroon & 5\\
        4 & Australia & 7\\
        5 & France & 6\\
        6 & Canada & 5\\
        7 & USA & 10\\
        8 & USA & 3\\
        9 & USA & 2\\
        10 & USA & 4\\
        \bottomrule
    \end{tabular}
    \caption{\newText{Interview participant demographics: their country of residence and number of years being involved in systematic counterspeaking.}{R3}}
    \label{tab:interview_demographics}
\end{table}

\subsection{Interview Study}
We first conducted semi-structured interviews with experienced counterspeakers to understand counterspeech from participants with diverse experiences countering hate and a more developed identity as counterspeakers. 
We asked questions to understand their counterspeech experiences and thoughts on AI tools, possible benefits and drawbacks, to provide insights into each of our research questions.

\subsubsection{Interview Procedure}
To gain an in-depth understanding of the challenges counterspeakers face, the strategies they use, and their thoughts about using AI to improve their counterspeech, we employed a qualitative research design, utilizing semi-structured interviews as the primary data collection method. 
We chose semi-structured interviews as they allow participants to express their thoughts, experiences, and perspectives, while also providing the flexibility to probe for deeper insights as the conversation unfolds. 
To ensure consistency, we developed an interview guide which consisted of open-ended questions designed to explore participants' experiences doing counterspeech (e.g., methods of finding hate speech, frequency and audience of counterspeech, and most rewarding experiences) and their insights into how AI tools could aid or complicate their work (e.g., their thoughts on counterspeech bots, open questions about envisioned AI tools). 
One of the authors conducted all interviews over online video calls in English between May and July of 2023. 

\subsubsection{Participant Recruitment}
Ten participants were purposefully selected based on their long-term experience of counterspeaking online in a systematic way. 
\newText{Recruitment was carried out by direct invitations through the partnered NGO.}{R3} 
The participants came from a variety of backgrounds in different contexts (from Europe, Africa, and North America).  
Some did counterspeaking collectively, while others responded individually. 
This sampling approach aimed to ensure a diverse range of perspectives and rich data for analysis. 
All participants provided informed consent prior to their participation, and confidentiality and anonymity were maintained throughout the study, with pseudonyms used in reporting findings.
\newText{We report their country of residence and years of experience counterspeaking in Table~\ref{tab:interview_demographics}.}{R3}

\subsection{Survey Study}

\begin{table*}[!ht]
    \centering
    \footnotesize
    \begin{tabularx}{\textwidth}{llX}
        \toprule
        Question Topic & QId & Question\\
        \hline
        \multirow{5}{*}{Hate Speech} & SQ1 & Which social media platforms or online spaces do you use at least once a week\\
        & SQ2 & How often, if ever, do you encounter speech online that you consider to be hateful?\\
        & SQ3 & On which social media platforms or online spaces do you feel like you see hateful speech most often (choose up to three)\\
        & SQ4 & \hangindent=1em Do you see more hateful speech in private online spaces (e.g., direct messages, private Facebook group) or public online spaces?	\\
        & SQ5 & Which of the following categories of hateful speech do you see most frequently? (select all that apply) \\
        \hline
        \multirow{7}{*}{Counterspeech} & SQ6 & \hangindent=1em How frequently, if ever, have you responded to hateful speech in a way that tried to counter the speech? (e.g., writing a denouncing or disagreeing comment, sending a private message, adding a negative reaction)	\\
        & SQ7 & How do you usually try to counter hateful speech? (choose up to three)\\
        & SQ8 & Which statement do you agree with most?\\
        & SQ9 & Who do you primarily respond to?	\\
        & SQ10 & \hangindent=1em  If you have written a reply to hateful speech online before, which of the following tactics have you used (check all that apply)\\
        & SQ11 & \hangindent=1em If you had to choose your most used tactic from that list, which would you choose (same question as before, but this time just tell us your most used tactic):\\
        & SQ12 & \hangindent=1em If you have seen hateful speech online before and chosen NOT to respond (react or write a reply), which of the following do you agree with? (choose all that apply)\\
        \hline
        \multirow{5}{*}{AI Tools} & SQ13 & \hangindent=1em If there was an AI tool to help you respond to hateful speech by specifically addressing the concerns you selected previously, how likely would you be to consider using it?\\
        & SQ14* & What are the reasons that you are likely or unlikely to use it?	\\
        & SQ15* & What type of AI assistance do you think would make you more likely to write a reply to hate speech?	\\
        & SQ16 & Select the following possible AI tools that would be useful for you to post a reply to hate speech. (select all that apply) \\
        & SQ17 & How do you feel about the following bots that automatically engage with hate speech? \\
        \bottomrule
    \end{tabularx}
    \caption{Questions listed in the public survey with participants on MTurk. * denotes an open-ended survey question.}
    \label{tab:survey-questions}
\end{table*}

To understand a broader user experience with hate speech and counterspeech, we conducted a survey study with participants from Amazon Mechanical Turk (MTurk). 
To ensure the quality of our results \cite{kennedy_clifford_burleigh_waggoner_jewell_winter_2020}, we pre-qualified workers using the process detailed in Appendix~\ref{app:qual} and excluded survey results with nonsensical qualitative answers (e.g., repetitive answers, discussing irrelevant technology, or using copy-pasted answers from other internet sources). 
The study was approved by our institution's IRB, and all survey responses were collected in August 2023.
The survey took median 8 minutes, and we compensated workers at a rate of \$12/hr.

\subsubsection{Survey Questions}
To get an overview of counterspeech experiences of a wider population, we designed a survey asking participants questions shown in Table~\ref{tab:survey-questions}. 
The three main parts of the survey were hate speech experience, counterspeech experience, and AI tools for counterspeech, followed by questions on demographics.
In the first part, we asked the participants questions about social media usage and their experiences with hate speech online such as types of hate and details about their experience (e.g., platform and type of online space - public or private) (5 questions).
The second part of the survey consisted of questions about counterspeech experience as previous responses to hate speech and barriers to responding (7 questions). 
The third part of the survey focused on questions about AI tools (e.g., openness to using an AI tool, preferences towards specific types of AI assistance) (5 questions). 
To avoid priming the participants toward specific type of responses on envisioned AI tools, we showed open-ended questions (SQ14 and 15) on separate screen to the survey questions that mention tools suggested by the research team.
Additionally, questions asking about hate speech and counterspeech experiences (SQ3-SQ14) were skipped for those who responded that they have never seen hate speech online (SQ2, option "Never", \textit{N=12}).

\subsubsection{Participant Demographics}
\label{sssec:surveydemographics}
\begin{figure*}[htpb]
  \centering
  \subcaptionbox{Hate speech (SQ2, \textit{N=342}) and counterspeech (SQ6, \textit{N=330}) experiences of participants shown through responses to likert scale question on frequency of hate speech encounter and counter hate speech frequency.\label{subfig:hate-counter-freq}}{%
     \includegraphics[width=0.8\textwidth]{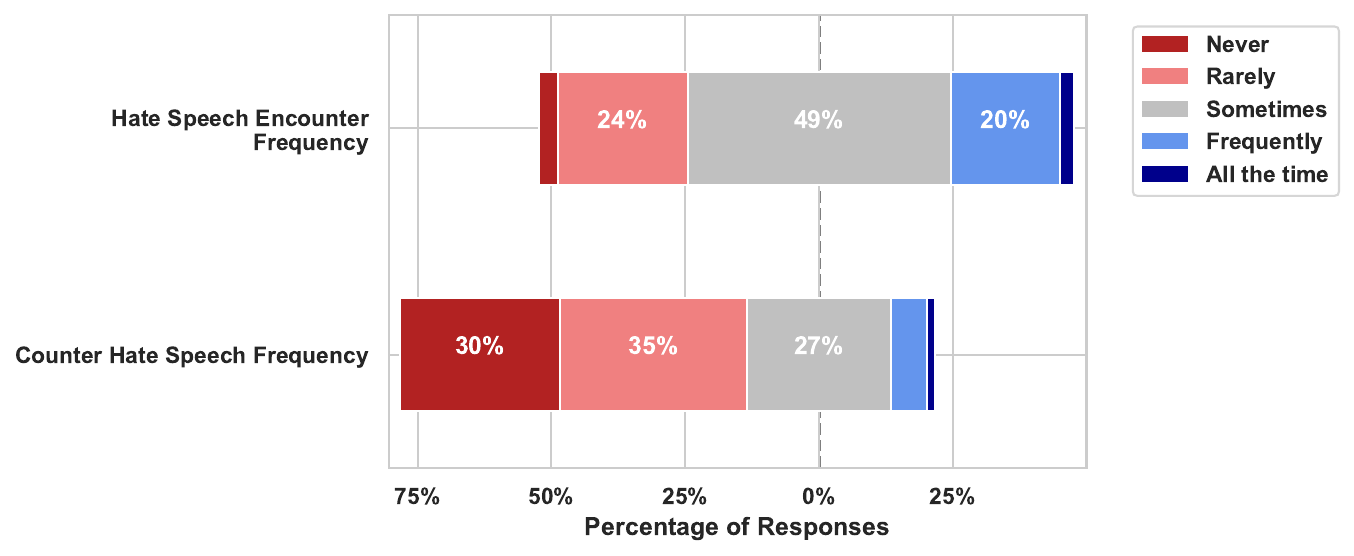}%
  }\hfill
  \subcaptionbox{Most commonly seen hate speech categories (SQ5, \textit{N=330}) identified by the participants shown in percentages of participants who selected the option. Participants were asked to select all that apply.\label{subfig:hate-type}}{%
    \includegraphics[width=0.9\textwidth]{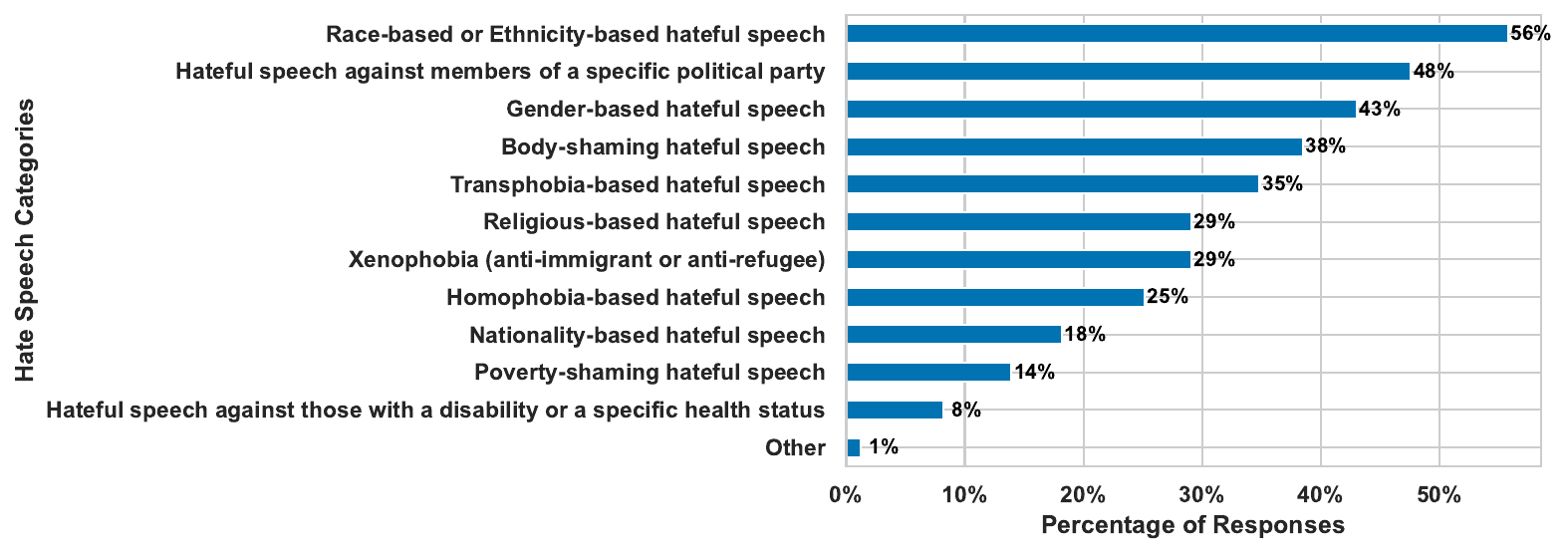}
  }
  \caption{Participant responses on hate speech and counterspeech experience questions and most commonly seen type of hate.}
  \label{fig:hs-cs-experiences}
\end{figure*}

Since we recruited mainly from U.S. and Canada, majority of the participants were from the U.S. (98\%) and many of them identified as white or Caucasian (83\%).
On question about gender identity, 56\% of participants identified as male and 41\% female.
When asked about sexuality, 85\% reported as being straight (heterosexual) followed by bisexual (6\%) and asexual (2\%).
On the political spectrum, participants were liberal-leaning with strongly liberal 22\%, liberal 32\%, moderate 18\%, conservative 17\%, and strongly conservative 8\%.
As shown in Figure~\ref{subfig:hate-counter-freq}, 94\% of the respondents had experience encountering hate speech online and 70\% had experience responding to hate speech (e.g., writing a comment, sending a private message, or adding a negative reaction) but only 8\% did so frequently or all the time even though 22\% encountered hate speech frequently or all the time. 
Out of those who had experience countering hate speech, 72\% had previously responded by adding a comment or reply under the post.
Moreover, the most commonly seen type of hate speech was race-based or ethnicity-based (56\%) (Figure~\ref{subfig:hate-type}), and most participants primarily responded either equally as an ally and targeted group (45\%) or as an ally (36\%).
Additional analysis of demographics of participants are shown in Appendix~\ref{app:dem}

\newText{
\subsection{Analysis Methods}
To conduct a comprehensive analysis, two separate groups of authors first performed grounded theory and open coding on the free text interview and survey data, respectively.
The authors then consolidated and synthesized the findings to build a cohesive theoretical framework.
Additionally, the quantitative responses were analyzed and integrated with the qualitative results.

\subsubsection{Data Preparation}
Interview data was transcribed by the author who conducted the interview to ensure participant confidentiality.
Moreover, the qualitative responses of the survey data were aggregated using quantitative methods to answer specific research questions. 
For barriers to counterspeech (RQ1, Section~\ref{ssec:barriers}), we looked at the open-ended responses to the survey question about barriers (SQ12, option "other"). For the possibilities of AI tools in counterspeech (RQ2, Section~\ref{ssec:rq2}), we analyzed the reason why participants were willing to use AI tools in counterspeech (SQ14) from participants who were willing to adopt AI tools in counterspeech (SQ12, options "likely" or "neutral"). We also analyzed participants responses on envisioned AI tools in counterspeech (SQ15). For the concerns of AI involvement in counterspeech (RQ3, Section~\ref{ssec:rq3}), we analyzed the reason why participants did not want to use AI tools in counterspeech (SQ14) from participants who were unwilling to use these tools (SQ12, options "neutral" and "unlikely").

\subsubsection{Qualitative Analysis}
\label{sssec:interview-analysis-method}
Grounded theory methodology \cite{charmaz2006constructing,glaser1967discovery} was employed to analyze the interview data. 
This iterative and systematic approach allowed for the discovery and development of theories directly from the data. 
The analysis process involved three key coding stages: open coding (i.e., generating initial codes by breaking down the data into meaningful units), axial coding (i.e., identifying broader categories and subcategories to establish relationships between codes), and selective coding (i.e., developing a theoretical framework by refining and integrating categories). 
Survey data was also analyzed using open coding but did not employ the last two stages of the grounded theory. 

\paragraph{Interviews}
The first two stages of coding were conducted by the same author who conducted the interviews, as was required by our IRB to protect participant confidentiality. 
During open coding, the interview transcript was broken down into meaningful units through line-by-line coding.
Then, stage three coding was conducted by the research team, integrating the findings of both the survey research and the interview study. 

\paragraph{Survey}
For qualitative survey data, two authors inductively generated codes for 25\% of the data by developing individual codebooks and labeling each instance with one or more codes. 
Each code was given a name and a short description. 
Next, the authors convened to merge their codebooks by combining codes with similar themes into a single code and unanimously agreeing to add, merge, or delete the codes through discussion in case of disagreement.
Finally, the authors performed a second round of coding by deductively applying the shared codebook to the entire dataset individually. 
The authors then reconvened and for each instance, applied the codes upon unanimous agreement based on discussion. 
Disagreements largely occurred due to differing scopes of codes and at times different interpretations of the statements, which were resolved through discussion.

\subsubsection{Quantitative Analysis}
\label{sssec:survey-analysis-method}
We report the quantitative responses with percentage of participant responses. 
For questions that allowed multiple choices (e.g., select all that apply, choose up to three), we calculated the percentage over the number of participants who responded.
Following best practices for opinion surveys~\cite{kitchenham2008personal}, we also aggregate similar responses together (e.g., extremely likely, likely).}{2AC\&R3} 

\section{Findings}
\label{sec:findings}

\begin{figure*}[!ht]
    \centering
    \includegraphics[width=\textwidth]{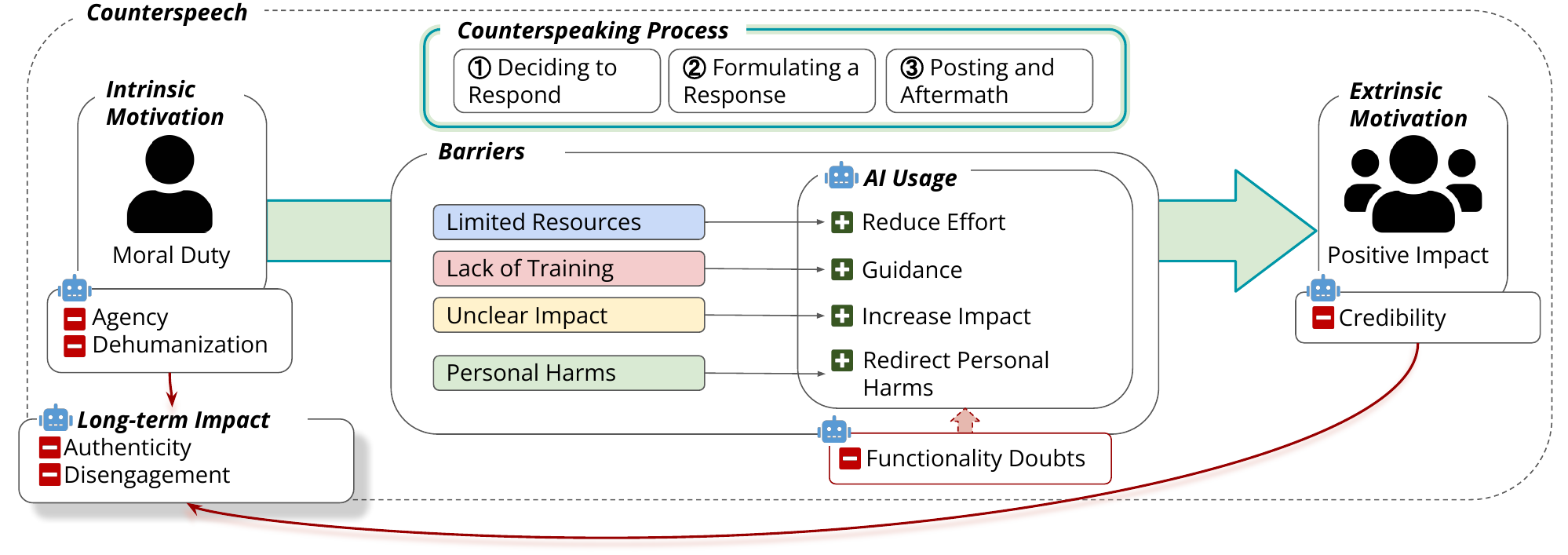}
    \caption{An overview of interactions between the themes surfaced in Section~\ref{sec:findings}. The figure, from left to right, show an overall counterspeech experience surfaced from participant responses. Participants' intrinsic and extrinsic motivation encouraged participants to engage in the counterspeaking process broken down into three steps. Themes found in beneficial AI usage {\color{plus}\faPlusSquare} could be rooted in each barrier, shown by the arrows in the figure, and themes found in AI concerns {\color{minus}\faMinusSquare} were linked to different aspects of counterspeech including motivations, functionality of AI for counterspeech, and counterspeech as a whole.}
    \label{fig:theme-interaction}
\end{figure*}

To understand counterspeech and potential impact of AI involvement from both activist- and lay-user perspective, we present findings from the analysis of both studies together in this section.
As shown in Figure~\ref{fig:theme-interaction}, we found that AI usage benefits could be mapped to the barrier each addressed.
Moreover, the concerns of AI tools could also be linked to the themes discussed in other section in detracting from intrinsic and extrinsic motivations, questioning the usage benefits, and negatively impacting counterspeech as a whole.

We first describe our theory of the counterspeech process, and then findings related to each research question: barriers to counterspeech (RQ1), possibilities of AI tools (RQ2), and concerns of AI involvement in counterspeech (RQ3). 
We denote interview participants as IP and survey participants as SP throughout the section for clarity. 
We highlight the names of codes by using an \textsf{\myuline{underlined font style}} throughout this section for visibility and clarity of connection between the findings and the code.
Codes developed for each section are listed in Appendix~\ref{app:analysis}.



\subsection*{An Inductive Theory of the Counterspeaking Process and Motivation}
\refstepcounter{theory}
\label{ssec:theory-of-counterspeaking}
In our analyses, we found that many counterspeakers shared a similar process for counterspeaking, and had similar themes for their motivation to engage.

\paragraph{Three-step Counterspeaking Process}
Our analyses highlighted that these three steps 
were commonly discussed in participants' counterspeaking process: (1) deciding to respond, (2) formulating a response, and (3) posting as well as engaging with the reactions of the audience. 
As a first step, counterspeakers either came across or actively looked for hate speech, assessed its harms, and made decisions based on effort and impact trade-offs on whether \textit{``it would be effective enough to be worth my time and effort''} (SP135).
After deciding to respond, participants formulated a response, usually comments or posts, to counter hate.
Some words used by participants characterizing the responses they would create were \textit{``thoughtful''} (SP331), \textit{``impactful''} (SP81), and \textit{``mindful''} (SP289).
The last step of counterspeaking was posting and dealing with the reactions including positive reactions such as \textit{``liking a comment or responding to a comment''} (IP9), negative \textit{``push back''} (IP4), or no reaction.
Each step of counterspeech required effort and time, although varying in amount, and a set of barriers existed, which are discussed further in Section~\ref{ssec:barriers}.

\paragraph{Counterspeaker Motivations}
Experienced counterspeakers expressed both intrinsic and extrinsic motivations, \textit{moral duty} and \textit{positive impact}, that encouraged them to take the above steps to engage against hate.

Counterspeakers we interviewed largely saw counterspeaking as a moral duty.
Many believed that it was the \textit{``\textup{\textsf{\myuline{right}}} thing to do''} (IP1).
Many found meaning in counterspeech in shared values as IP7 noted:
\begin{quote}
    \textit{``I believe in it. I didn't do it just because I got a kick out of it.''}
\end{quote}
The sense of moral duty towards counterspeech was also shared by survey participants as SP107 wrote:
\begin{quote}
    \textit{
    ``I would really like to be able to make the internet a safer and more healthy place to spend their time, so if I could reduce the amount of hatred and misinformation on there, I would do it.''}
\end{quote}
However, some survey participants disagreed with this perspective sharing that \textit{``I did not respond because for the most part people should be able to say what they want. It's up to us whether we choose to be hurt by someone's comments or not''} (SP190). 

Another extrinsic motivation discussed by many participants was the positive \textsf{\myuline{impact}} of counterspeech. 
The types of impact participants found most rewarding varied, ranging from influencing the conversation to causing a view change or consoling those targeted by hateful speech. 
For example, participants mentioned the following rewarding experiences: seeing a positive change in the comment section, learning that their counterspeech \textit{``changed the mind of (even) one person''} (IP1), and knowing that they have \textit{``helped someone who had maybe been reading the comments and had been upset by them''} (IP4).
Participants emphasized that they felt rewarded even when the scale of their impact was small.

\subsection{Counterspeech Barriers (RQ1)}
\label{ssec:barriers}
To answer our first research question (RQ1), we analyzed participant responses around pain points of counterspeech and reasons behind not engaging in counterspeech.
As shown in Figure~\ref{fig:theme-interaction}, there were four high-level themes in counterspeech barriers identified by the participants: \textit{limited resources}, \textit{lack of training}, \textit{unclear impact}, and \textit{personal harms}.

The barriers that were discussed commonly across all stages were required \textsf{\myuline{resources}} such as time and energy and personal harm on \textsf{\myuline{mental health}}.
Overall, participants discussed that counterspeech can \textit{``take a lot of resources''} (IP2) and can be overwhelming as \textit{``it sometimes gets to be too much''} (IP5).
Since most interview participants were volunteer counterspeakers, there was no compensation for their time or harm to their mental health. As IP2 shared, \textit{``No one pays me - it's the time investment''}.
Similarly, the resource barrier was reflected in the survey responses as 21\% of the participants answered that they did not write a reply because they did not have enough time (Table~\ref{tab:cs-barrier}). 
The concern for mental health was also shared. As SP12 said, \textit{``I would get too upset about it and I have enough stress already in my life.''}
For the remaining discussion of counter speech barriers, we organize our discussion of how these barriers occur with respect to the three steps of the counterspeaking process.

\begin{table*}[htpb]
    \centering
    \small
    \begin{tabularx}{\textwidth}{Xl}
        \toprule
        \rowcolor{pastelgray}\cellcolor{pastelgray}\textit{\textbf{Counterspeech Barrier}} & \cellcolor{pastelgray}\textit{\textbf{Response, \% (\textit{N=330})}} \\
        I did not respond because I didn't think responding would have an impact & 73 \\
        I did not respond because I didn’t want people to be mad at me or send me hateful or threatening messages & 31 \\
        I did not write a reply because I did not know what to say & 22 \\
        I did not write a reply because I didn’t have enough time & 21 \\
        I did not write a reply because I didn't know how to say what I wanted to say & 17 \\
        Other & 9 \\
        \bottomrule
    \end{tabularx}
    \caption{Options and participant responses to question about counterspeech barriers (SQ12 in Table~\ref{tab:survey-questions}). Participant responses are shown in percentages of participants who agreed with the listed reason for not responding. Participants were asked to choose all that apply.}
    \label{tab:cs-barrier}
\end{table*}


\paragraph{Deciding to Respond}
Our findings showed that the most influential consideration to the decision of counterspeech was in the resource and impact trade-offs.
As a way of strategizing for impact, activists spent time \textsf{\myuline{finding hate speech}} looking for interactions where it would be \textit{``worth it''} (IP1) for them to respond.
They often looked at not only the hateful post but also the interactions around it (e.g., activity level of the comment thread), which could take additional time as reflected in IP10's experience of spending \textit{``up to two hours looking for good actions''}.
Similar effort to assess resource and impact trade-offs had to be made in discerning whether the hate speech was coming from \textit{``trolls''} (SP115) or \textit{``bots''} (SP341).
For example, SP232 shared, \textit{``I didn't know if the person who stated is in [sic] actor paid to say it so my response would be meaningless or that the writer is actually just a bot''}. 
Correct assessment of the impact was further complicated by algorithms on the platform as noticed by SP227, \textit{``I don't want to engage \& help the comment have more impact''}.
Supporting these decision considerations shared by the participants, the unclear \textsf{\myuline{reach}} of counterspeech at this stage was the most influential reason to not engage (73\%; see Table~\ref{tab:cs-barrier}). 

\paragraph{Formulating a Response}
Participants shared that this stage can be one of the most time-intensive as IP6 reflected, \textit{``It takes me forever to craft something that would make sense.''}
Moreover, lack of \textsf{\myuline{training}} was also noted as a barrier at this stage as people \textit{``don't always know what to say''} (IP3) to \textsf{\myuline{reach}} their audience to effectively counter hate. 
This can be especially true for beginners. 
As IP7 recalled, when he first started counterspeaking, he was \textit{``not as well-versed in the subject or in framing the argument and being persuasive.''}
Survey results also showed that lack of training was a barrier to counterspeech, however, not all participants responded by formulating their own message. 
Out of those who had responded to hate speech before, 72\% of the survey participants said they had previously responded by commenting or posting a reply, but some chose to rather use existing response methods such as downvoting or disliking the post (70\%) or reporting the hateful post (49\%).
While not all respondents had experience writing counterspeech, when asked about reasons for not engaging, 22\% of participants did not know what to say and 17\% did not know how to express what they wanted to say (see Table~\ref{tab:cs-barrier}).
Survey participants used various tactics in their responses.
Most commonly, they had ``tried to correct misinformation or fact-check inaccuracies'' (70\%) or ``tried to shame the person who has posted hateful speech'' (36\%) but also posted links to other sources (34\%), tried to be funny (21\%) or emotionally connect (21\%).

\paragraph{Posting and Aftermath}
At this step of the counterspeaking process, the lack of reactions and negative reactions are barriers to future action.
Counterspeaker activists reported being discouraged or demotivated by the limiting \textsf{\myuline{reach}} as one participant highlighted, \textit{``When you feel unheard and it’s like I’m doing this for nothing - it’s not really getting the word out - it’s frustrating''} (IP1).
Counterspeech sometimes caused participants to become the target of hate as well, negatively impacting their \textsf{\myuline{mental health}}.
For example, IP4 recalled that \textit{``For a while, the push back was getting to me''}. 
Therefore, counterspeakers opened themselves up to varying amounts of \textsf{\myuline{risk}} when posting, and it became a significant barrier as shared by IP4, \textit{``The less I say, the less room I give people to attack.''}
This is similarly reflected in the public survey opinion, as 31\% of participants saying they did not respond because they did not want to upset others or receive hateful messages (see Table~\ref{tab:cs-barrier}).
Additionally, survey participants saw more hate speech in public online spaces (84\%) and were primarily responding to audiences that included people that they did not know (94\%).
As IP1 reflected, \textit{``it’s risky when you put yourself out there''}, highlighting that responding to hate speech can open up confrontation in public spaces with strangers. This risks could be amplified when counterspeakers commented about highly controversial topics or did their work while living under authoritarian regimes or in conflict zones. 


\begin{figure*}[htpb]
  \centering
  \subcaptionbox{AI tool perception (SQ13, \textit{N=330}) shown in percentages of responses to each likert scale option.\label{subfig:ai-tool}}{%
     \includegraphics[width=0.8\textwidth]{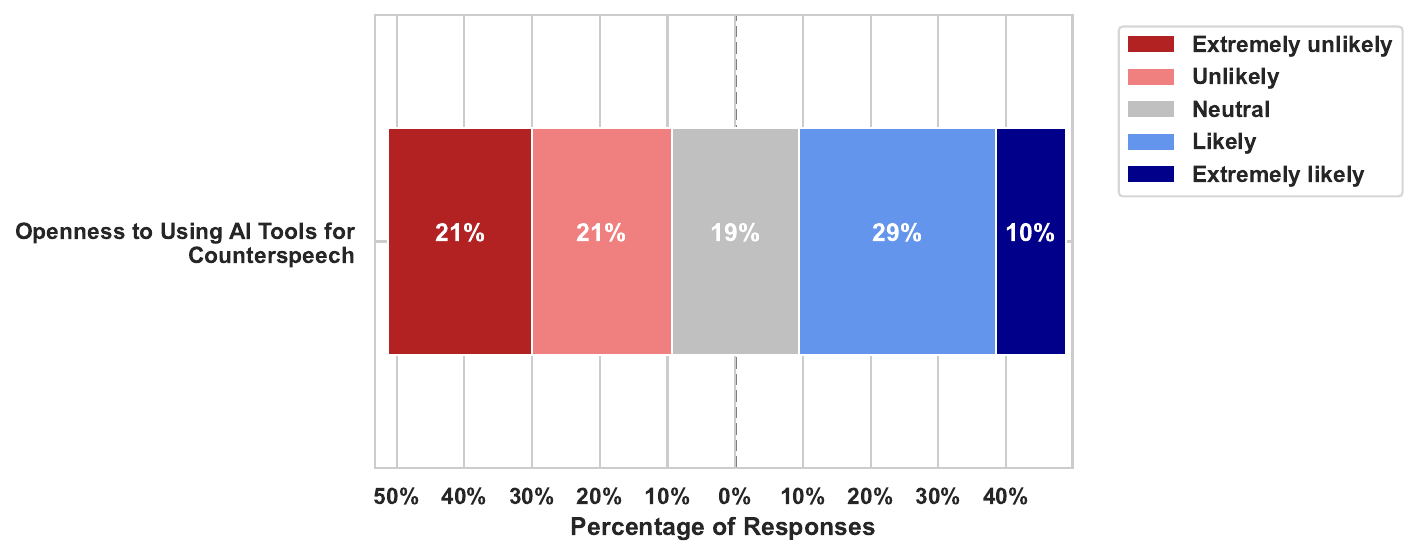}%
  }\hfill
  \subcaptionbox{Collaborative tool preferences (SQ16, \textit{N=342}) showing the percentage of participants who selected the tool as potentially useful. Participants were asked to select all that apply\label{subfig:tools}}{%
    \includegraphics[width=0.9\textwidth]{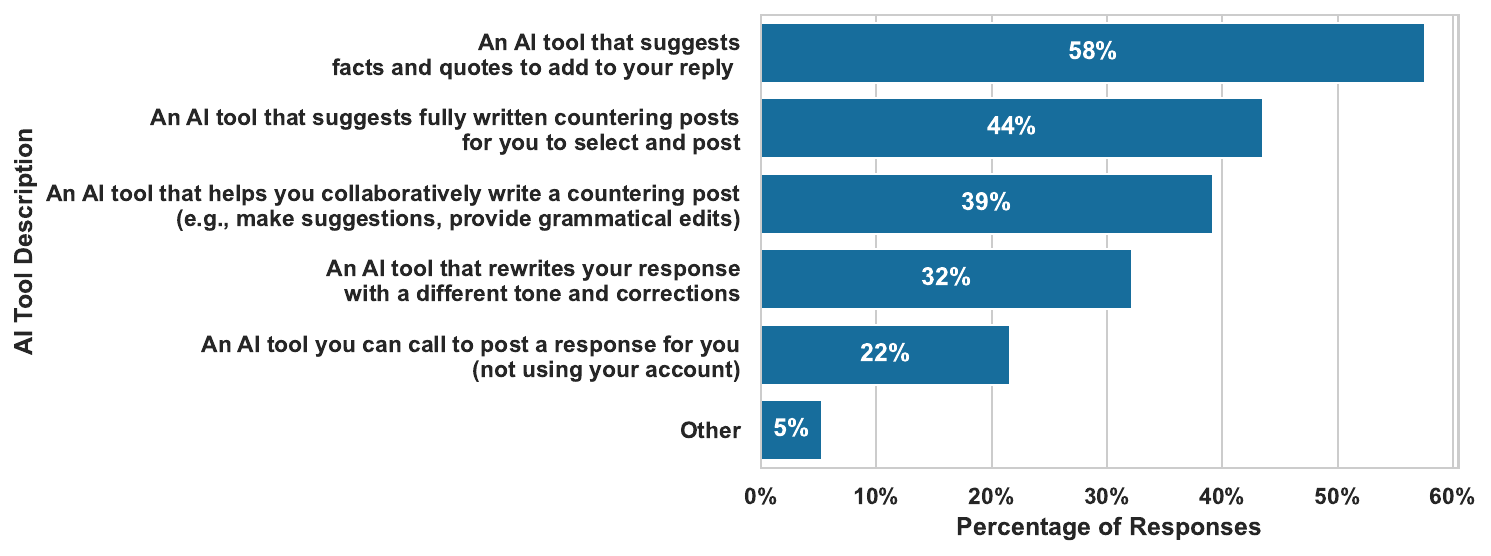}
  }\hfill
  \subcaptionbox{Bot preferences (SQ17, \textit{N=342}) shown in percentage of responses to each likert scale option\label{subfig:bots}}{%
    \includegraphics[width=0.8\textwidth]{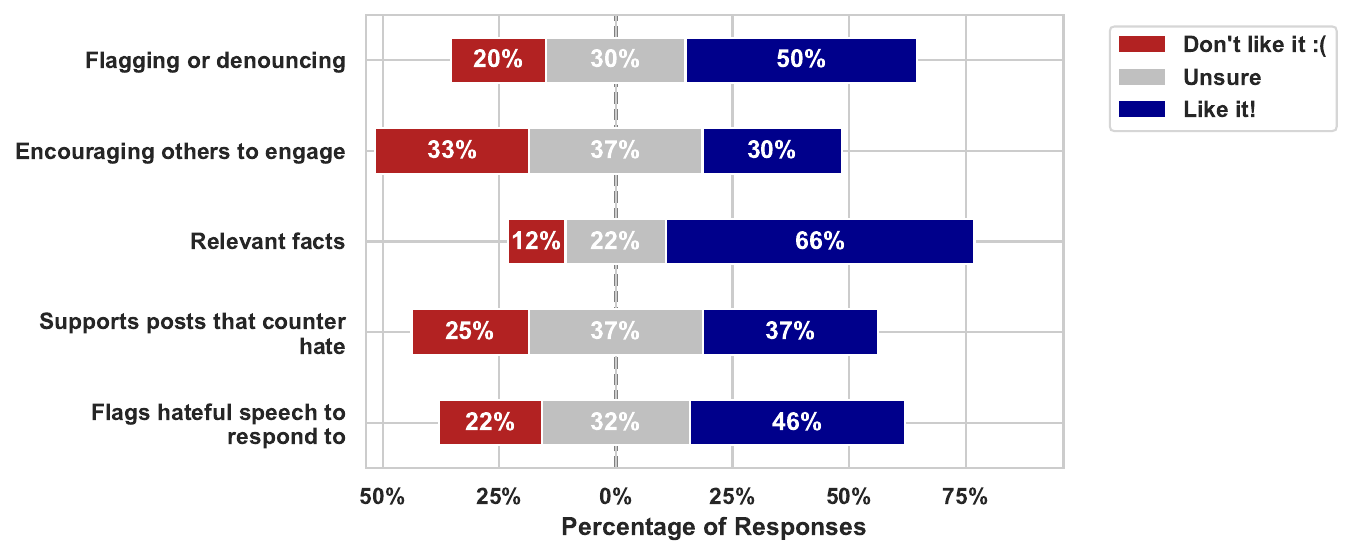}
  }
  \caption{Survey participant responses on quantitative questions about AI tools.}
  \label{a-two-column-figure}
\end{figure*}

\subsection{Possibilities of AI Tools (RQ2)} 
\label{ssec:rq2}
To answer our second research question (RQ2) on possible AI tools, we discuss our findings from the analysis on participants' responses about the benefits of AI tool usage and their description of tools that would encourage them to write a response.
Overall, the biggest difference seen between the two populations was in level of AI involvement; active counterspeakers saw possibilities for using AI to augment their work to make the process more efficient and amplify their voices, whereas lay-participants expressed a more diverse set of needs and preferences.
The usage goals and characteristics could be mapped to the themes of barriers they addressed as seen in Figure~\ref{fig:theme-interaction}. We break these down into a description of the types of support, empowerment, emotional, and factual support and the types of tools, level of involvement in the counterspeaking process and the definition of usability. 


\paragraph{Empowerment}
Some participants also characterized their preferred tool as empowering and aligned, with themselves or cultural societal norms, especially towards creating a positive impact.
For example, SP97 was interested in using AI tools \textit{``So that I can help people being attack(ed) by these hateful speech.''}
Similarly, SP139 showed positive interest towards AI tools \textit{``Because it might increase the chance to actually make an impact''}.
SP28 was interested in the support of AI tools \textit{``Because it would help me speak up more.''}
Participants also wanted tools that were \textsf{\myuline{aligned}} with them, which could \textit{``write a response EXACTLY LIKE I WOULD''} (SP30).
Some acknowledged that tools also need to be aligned with cultural and societal norms to create \textit{``unbiased''} (SP126) responses. 


\paragraph{Emotional Support}
Participants brought up the need for emotionally supportive AI tools in various ways.
On one hand, some participants thought that AI tools could help them with regulating their own emotions as they believed that AI's \textit{``emotional detachment could help me remain composed''} (SP160).
On the other hand, others thought that AI tools could help connect people to \textit{``think in a more empathetic way''} (SP262) focusing on different perceived AI capabilities and stylistic preferences.
Emotional support was also a commonly brought up need to help formulate an \textsf{\myuline{effective communication}}. Most notably, participants wanted tools to help communicate or understand emotions that shows emotional intelligence.
For example, SP107 proposed a tool with an \textit{``AI that could delve into the psychology of why a person posted what they did, and then give me a response that could really tap into that person's mind and give me a response that they will actually care about.''}

\paragraph{Factual Support}
Many participants were also interested in AI tools that could gather or show relevant \textsf{\myuline{information}}.
These participants wanted support in crafting \textsf{\myuline{factual}} and logical responses, either via resource recommendation or through fact- or argument-checking their own responses, to correct misinformation in the hateful speech.
An example of such tool, discussed by SP9, would \textit{``pull in articles to prove a point/correct misinformation''}.
In corroboration of these desired needs, fact-based support was the most preferred in our quantitative survey responses as evidenced by relevant facts bot (66\%) and an AI tool that suggests facts (58\%) being the most selected or liked AI usages (Figure~\ref{subfig:bots} and \ref{subfig:tools}).


\paragraph{Level of AI Involvement}
Participant responses showed preferences towards varying amount of AI involvement at different stages of the counterspeaking process (Section~\ref{ssec:theory-of-counterspeaking}).
In the first step of the process (\textit{deciding to respond}), some survey participants showed preference for an AI tool to automate this step as a part of a full automation, i.e., that detects hate speech and automatically \textsf{\myuline{reports}} and/or \textit{``blocks or deletes''} (SP148).
Some interview participants also expressed interest towards a detection tool to \textit{``identify hate speech more efficiently''} (IP5), however, unlike the survey participants, wanted full control of the subsequent counterspeaking steps. 


In the second step (\textit{formulating a response}), the varying levels of involvement ranged from \textsf{\myuline{collaborative}} tools to provide \textsf{\myuline{guidance}} \textit{``to find arguments to counter hate speech more effectively''} (SP104) to response support tools like \textit{``one(s) that recommends possible replies''} (SP132) .
The collaborative tools described by participants were characterized by less AI involvement with more input from users that would provide correction of \textit{``mistakes like grammatical mistakes, factual mistakes, etc.''} (SP256), \textit{``customization options''} to preserve communication style (SP160), or \textit{``help with idea generation''} (SP231).

At the last step of the counterspeaking process (\textit{posting and aftermath}), participants often wanted heavier AI involvement to reduce personal harms.
One type of support identified was to provide an \textsf{\myuline{AI proxy}}, often in anonymity, to counterspeak \textit{``without worrying about potential blowback''} (SP48).
Moreover, participants wanted tools to help deal with possible negative reactions with a protective AI tool, for example, \textit{``An AI that would respond to the hateful things that people said back or messaged to me.''} (SP110).

Overall, participants wanted AI tool's involvement to also help reduce harm to mental health by making counterspeaking \textit{``less stressful''} (SP35).
Additionally, IP3 proposed an educational tool that could be helpful to address the barrier in lack of training, \textit{``It would train people and walk them through the process of how to design a campaign.''}
Notably, the two populations showed different levels of preferred AI involvement, as lay-participants endorsed full automation and showed more interest towards heavier involvement.


\paragraph{\newText{Defining Usability}{2AC}}
The most prevalent preference surfaced was towards a \textit{usable} tool.
\newText{While this result is unsurprising, understanding aspects of usability could be helpful as to avoid building tools that are confusing, difficult to learn, or even useless.}{2AC} 
Participants described usability in two dimensions, ease of adoption and functional. 
Especially to address the barriers related to limited resources, participants wanted \textsf{\myuline{efficiency}} and to save \textsf{\myuline{time}}, as specified by SP9 as \textit{``something that makes it as quick and easy as possible to reply''}.
Some participants further detailed a possible tool that would be \textit{``provided by the site itself so I would not have to waste time getting a reply from another site''} (SP95) highlighting that it should cause a minimal amount of disruption or overhead.
There were varying perceptions of the functionality of AI systems, as some survey participants believed that AI would be better than them in being \textit{``more effective''} (SP169) by having \textit{``more experience''} (SP201), whereas others were more skeptical and wanted to \textit{``try it to see if it is beneficial for me''} (SP6).
\newText{Thus, many participants echoed curiosity around usability of the systems, which depended on their functional benefit.}{2AC}

\subsection{Concerns of AI Involvement in Counterspeech (RQ3)}
\label{ssec:rq3}
Our final research question (RQ3) asked about concerns regarding AI involvement in counterspeech.
To answer this question, we present our findings from analysis of interview and survey responses that presented reasons against using AI tools or showed concerns about AI involvement. 
Our analysis surfaced four themes: short-term concerns of \textit{credibility}, long-term negative impact on \textit{authenticity} leading to \textit{disengagement} in counterspeech, loss of \textit{agency} over and \textit{dehumanization} of counterspeech, and \textit{functionality doubts} about AI tools. 
As illustrated in Figure~\ref{fig:theme-interaction}, concerns about AI involvement could be connected to their negative impact on motivations discussed in the Section~\ref{ssec:theory-of-counterspeaking}, benefits of AI tools mentioned in Section~\ref{ssec:rq2}, or counterspeaking as a whole in long-term.


\subsubsection{Loss of Authenticity and Agency}
\label{subsubsec:auth-n-agency}
Authenticity and agency were overarching concerns across the two different populations of participants. 
In the short term, AI involvement could make counterspeech seem insincere and reduce speaker credibility to the audience (e.g., original poster, bystanders).  
For more experienced counterspeakers who have built an audience and a rapport, using AI could pose a risk of losing this trust. 
For example IP7 said, \textit{``If people figure out it's a bot (an automated response), then it loses all credibility.''}
Similarly, sincerity of care about standing up against hate, especially in the choice to stand up for specific topics, made counterspeech meaningful as a moral action as expressed by SP294:
\begin{quote}
    \textit{``If I'm going to respond to hate speech, I want it to come from me, because it's something I stand up for.  I wouldn't want an AI to be apart [sic] of something so important to my personality and morality.''}
\end{quote}
Authentic care and intentions were also central to connection as shared by IP8:
\begin{quote}
    \textit{``The process itself I find very satisfying. Having a sense of not being alone. You have all these people from all over the world, and you consider them friends, which is crazy, because you don't know them. For me, what's really touching is that someone out there is just there to support me. I'm not the only one who thinks this.''}
\end{quote}
Thus, participants emphasized the value of credibility and sincerity in their work as shared by IP2 that \textit{``If you are trying to be genuine, then you have to, you know, be a genuine human.''}

Participants also shared concerns over \textsf{\myuline{agency}} in the counterspeaking process that might arise with the use AI tools.
For example, SP6 said, \textit{``I like using my words not being tied to what AI says''}.
Similarly, interview participants also communicated concerns of agency, reflecting that their identity as \textsf{\myuline{counterspeakers}} and more broadly as moral agents were reinforced through counterspeaking. 
For example, IP8 described how empowering it is when you make a visible impact after responding to online hatred:  
\begin{quote}
    \textit{``There is something incredibly magical about turning something really hateful in the other direction. You feel you aren't hopeless or helpless.''}
\end{quote}
Experienced counterspeakers also discussed their concerns about making moral compromises by using similar tools as those using hate speech bots making them \textit{``no better than what is being used''} (IP6).
Additionally, SP277 discussed possible moral degradation of counterspeech that would be made permissible by AI tools:
\begin{quote}
    \textit{``I would like to have the support of the tool, and to be honest it sort of makes me feel like I have some plausible deniability if an issue arises. In a worst case scenario I would be able to "blame" it on the AI.''}
\end{quote}

Participants worried that over time these concerns could develop into \textsf{\myuline{long-term}} negative impacts on online communication.
At a larger scale, excessive use of AI tools, especially without meaningful human oversight, could make joining \textsf{\myuline{real}} activism and finding genuine connection more difficult. 
IP5 warned against participation fatigue, comparing AI counterspeech to petitions:
\begin{quote}
    \textit{``Look at petitions. There was a moment when petitions were kind of rare. Now you are harassed with petitions, and they don't have any purpose anymore. They are creating a false, passive attitude that 'I've already done my bit.' Why would we want to replicate that?''}
\end{quote}
Moreover, without authentic intentions behind counterspeech, possible AI automation could reduce the meaning of counterspeech as IP6 emphasized through a comparison to robot fights:
\begin{quote}
    \textit{``Where is the human component to that? Yeah, like, it's like watching those robot fights where it's just the robots. It's like, I don't know, then it becomes a game, right? And it almost, I don't know if the word is dehumanizes, but it desensitizes people from what is actually going on.''}
\end{quote}

Thus, careless AI involvement could exacerbate an already prevalent attitude towards disengagement shared by lay-participants who believed that \textsf{\myuline{engaging was not helpful}} and would rather \textsf{\myuline{avoid}} hate speech because they did not care enough.
This sentiment was related in the response by SP137 who was not likely to use an AI tool because:
\begin{quote}
    \textit{``[It] just seems like a waste of time that will create the prospect of them using an AI tool to respond to me. This will result in both of our AI tools going back and forth indefinitely and won't solve anything overall.''}
\end{quote}

\subsubsection{Doubts in AI Capabilities}
\label{sssec:doubts}
\textsf{\myuline{Functionality doubts}} of whether AI tools can actually address these barriers were also shared between the two populations of participants. 
Despite some survey participants calling for emotionally aware AI tools (Section~\ref{ssec:rq2}), many participants believed that AI did not have the \textsf{\myuline{emotional intelligence}} to adequately counterspeak.
Some noted that they are not \textit{``funny or clever''} (IP2), and highlighting the lack of empathy in these systems, IP1 said, \textit{``It couldn't cover the human aspect of empathy''}.
Moreover, participants distrusted AI, sharing their perception that \textit{``AI tools are often wrong and I wouldn't want its bias to affect what I am posting''} (SP143).
The limitations of its training data and cultural bias were noted by IP3 who shared the concern that \textit{``These are trained on Western data, so I immediately found a problem with that... Hate speech in Cameroon is definitely not [the same as] hate speech in the United States.''}
AI assistance was also seen as limited in solving the problem of personal harm such as \textsf{\myuline{becoming the target}} of retaliation, and SP156 noted that AI involvement would not solve the platforms' algorithmic problems to counterspeech as it would still \textit{``drive(s) traffic to it (hate speech) which makes it a bigger problem''} (SP156).
\section{Discussion}

To answer how AI assistance should be developed to improve the process of counterspeaking and not to detract from its meaning, we conducted interviews and surveys with both experienced counterspeakers and everday social media users to investigate three research questions: what are barriers to counterspeaking (RQ1), how could AI tools assist in counterspeech (RQ2), and what are some concerns about AI involvement in counterspeech (RQ3).

Our analyses surfaced barriers and AI needs with four different themes to inform functional needs of AI assistance and found that many participants thought AI tool could be empowering. 
Moreover, we discovered themes of counterspeaker motivations, especially in connection to oneself and others, highlighting the human components of counterspeech. 
Through understanding participants' concerns of AI involvement, we identified that without careful considerations, AI tools could do more harm than good, detracting from counterspeaker motivations and reducing meaning in communication.  
Based on our findings we build a set of recommendations and considerations for designing beneficent AI tools for counterspeech.

\newText{
\subsection{Possible Tools to Address Barriers to Counterspeech}
Our study participants described many different barriers to counterspeech at various stages of engagement as discussed in Section~\ref{ssec:barriers}.
However, previous works in AI have focused on a narrow set of challenges for assistance and automation: automatic counterspeech generation \cite{Zhu2021-prune,halimWokeGPTImprovingCounterspeech2023,Saha2022-countergedi,gupta2023counterspeeches} and analysis and detection of both hate speech and counterspeech \cite{garland-etal-2020-countering,hassan2023discgen}.
Our findings paint a broader picture, especially through the theory of counterspeech engagement, and highlight where tools and resources would encourage bystander intervention towards countering hate. 
More specifically, participants described needs for further research and assistance in education, bot detection, and online safety.
Education and training in all three stages (e.g., learning when to engage, how to engage, and how to handle backlash) could simplify and encourage counterspeech. 
For example, classrooms can play an essential role in counterspeech education as teachers already play an important role in educating students in handling digital risks and experiences \cite{maqsood2021tween}.
However, as \citeauthor{Castellví2022} has shown, training and resources are needed to equip them with diverse strategies to construct effective counterspeech \cite{Castellví2022}.
Therefore, resources in counterspeech education, which could be enhanced by AI \cite{breideband2023cobi}, and continued work on understanding different strategies and effective responses \cite{mun2023denouncing,hangartner_empathy_2021} are highly necessary. 

Moreover, participants highlighted the importance of \textit{impact} in their decision to respond. 
In their consideration, knowing whether their comments would have an impact to their recipients or others mattered. 
Therefore, efforts in counterspeech research to understand and model the impact of counterspeech could be valuable to users to understand the impact of their actions, which could be aided by AI and algorithmic tools \cite{mathew2020begets,garland2022impact}.
Many participants also noted that they did not want to respond to a bot, and as previous works have shown, social bots (i.e., algorithms that produce content and interact with users on social media) are an ongoing problem in many social media platforms \cite{ferrara2016bots}. 
AI bot detection using neural methods have been explored to help social media platforms in filtering bots \cite{Yang2019ScalableAG,cresci2020botdetection}, but with increasing capabilities of generative AI, detecting non-human agents is becoming harder \cite{ghosal2023possibilities}.
Therefore, both legal efforts to deter bot accounts \cite{strickePeopleRobotsRoadmap2019} and research on bot and AI detection would have to progress together to reduce the uncertainty and noise in online communication. 

Another area of support highlighted by the participants was in online safety to overcome the challenges in posting and aftermath. 
In a survey conducted in 2017 and again in 2022, 41\% of Americans reported experiencing online harassment \cite{vogels2021state} supporting the fear of retaliation experienced by bystanders \cite{wong-loDigitalMetamorphosisExamination2014} as related by our participants.
The risk of intervening against online hate speech can also be exacerbated by gender roles and 
other identities of the intervener \cite{tianyi2023intervention,wilhelmGenderedMoralityBacklash2019}.
While AI and algorithmic tools for moderation and flagging have been developed to alleviate online harassment and burden of moderators \cite{blackwellClassificationItsConsequences2017}, they have shown to have varying levels of effectiveness \cite{sap-etal-2022-annotators} especially for marginalized communities who perceive greater harms associated with online harassment \cite{imWomenPerspectivesHarm2022}.
Along with improving existing methods and tools, further support in harassment reporting (e.g., collecting evidence or adding context) and recovery support could be assisted by AI-enhanced tools \cite{goyal2022harassment}.
Moreover, policies from social media platforms and legal authorities have varying and unclear definition and responses to online harassment \cite{paterCharacterizationsOnlineHarassment2016} which are often not reflective of cultural contexts \cite{schoenebeck2023harassment}.
This could cause interventions from users and AI tools to be confusing and ineffective.
Therefore, continued investigation of online harassment through both research and policy efforts is necessary to create safer online spaces, which would not only encourage intervention from bystanders by minimizing retaliation harms to counterspeakers but also help define the role of counterspeech in this effort.
}{2AC}

\subsection{Preserving Authenticity of Counterspeech}
One of the overarching concerns of AI tools that participants raised was authenticity of counterspeech and transparency of online communication.
Participants raised concerns that AI involvement would cause counterspeech to be viewed as insincere or less credible.
Moreover, many participants pointed out that AI tools or bots are frequently used to generate hate speech (Section~\ref{subsubsec:auth-n-agency}), thus leading to the idea that automated AI counterspeech would be pitting bots against bots (e.g., \textit{protesting bots}).
Previous work supports this need for authenticity when speaking up against hate, for example in creating solidarity in activist movements and in calls to bystanders to organize and participate \cite{horowitz2017we,SCHULTE2020identity}.
This concern is also echoed by literature on AI which has shown that use of AI in communication can negatively affect trustworthiness of the speaker and authenticity of the message \cite{jackesch2019aimc,gilkson2023aimcaplogy}. 

Authenticity and trust is, therefore, a key consideration for designing AI tools that can help counterspeakers without damaging their message. 
To help promote authenticity in online communication and reduce information overload due to unclear sources \cite{laato2020unverified}, any system that generates automated counterspeech responses (bots) should clearly disclose that it is not a human \cite{etzioniOpinionHowRegulate2017}. 
Bot and AI generated text detection \cite{cresci2020botdetection,sadasivan2023aigenerated}, and disclosure policies \cite{Weaver2018} will also help ensure transparency in online communication. 
For responses generated through human-AI collaboration, there is less clarity on how much disclosure is necessary to aid authentic communication and connection. 
Other AI-mediated communication tools have been shown to create different perception of the writers and their intentions and led to feelings of deception \cite{hancock2020aimc,Hohenstein2023-zj}.
AI counterspeech tool, especially because of its closeness to morality (Section~\ref{ssec:theory-of-counterspeaking}), may further exacerbate these negative effects. 
Therefore, it is an important research direction to understand how to frame and present communication that is collaboratively created with AI.
\subsection{Cultivating Moral Agency and Engagement}
Many participants saw counterspeech as a moral imperative believing that it was the right thing to do and chose to speak up against hate in ways that reflected their values (Section~\ref{ssec:theory-of-counterspeaking}).
This is consistent with the argument by \citet{Howard2021-ll} that counterspeech is a moral duty for all citizens.
However, with AI involvement, some participants showed concerns about creating passive moral attitudes leading to disengagement or shifting responsibilities to AI, echoing the concerns of moral passivity caused by AI \cite{Danaher2019patiency} and misattribution of responsibility in AI mediated communication using AI as a scapegoat when conversations go awry \cite{HOHENSTEIN2020aimcmoral}.

Therefore, cultivating moral agency is an important design consideration to build AI tools that empowers users to engage and take responsibility towards moral duties. 
Future research on AI assisted counterspeech should focus on guiding users to be deliberate in their choices to engage and to not overrely on AI systems to make moral choices.
Design methods to reduce overreliance such as cognitive forcing function such as checklists \cite{bucina2021think} or explanations about its outputs \cite{vasconcelos2023explanation} should be explored and integrated into design of such AI systems.
Furthermore, transparency about failures of AI systems, notably encoded biases \cite{weidinger2021llmharm,farina2023gptissue} and limited cultural context \cite{santy-etal-2023-nlpositionality}, would also be an important feature as AI generated text can instill values and norms \cite{Krugel2023-sg,Jakesch_2023}.
In addition to building unbiased and culturally aware AI systems \cite{mehrabi2022bias}, customization could help promote user agency and correct these shortcomings \cite{kang2022agency,usmani2023ethical}.
However, customization can lead to more cognitive effort, so future AI counterspeech tools should consider effort-agency tradeoffs and explore distinctions between meaningful and non-meaningful effort \cite{INZLICHT2022effort}.

\subsection{Protecting Mental Health}
The stress of responding and not caring enough were frequently discussed as reasons behind participants' choices to not engage. 
Experienced counterspeakers also noted that constant exposure to hate speech can lead to feelings of being overwhelmed and it becoming too much (Section~\ref{ssec:barriers}).
This is consistent with literature on the experience of cyberbullying victimization, becoming the target of cyberharassment or hate speech, and its link to mental health, especially in relation to depression, anxiety, and social media fatigue \cite{schodt2021cyber,cao2019smf,JENARO2018cyber}.
\newText{Exposure to hate speech can especially be damaging for the minority groups being targeted and can lead to depression, suicidal thoughts, and stress \cite{zochniak2023wellbeing}.}{R3}
Some recipients of counterspeech (i.e., speaker behind hate speech) are not intentionally saying harmful things and could be willing to change. 
Therefore, call-out based \cite{ross2019speaking} online shaming \cite{Thomason2021shame} and domination \cite{whitten2023republican} which could escalate conflict or lead to harm \cite{Lepoutre2022} should be avoided. 

Thus, any AI counterspeech tool should take into consideration mental health of those involved to empower and not harm users.
One recommendation would be to design tools focusing on mental health such as human-AI collaboration for empathetic conversations \cite{Sharma2023} and focus on guiding systems to reflect a call-in culture that encourages relating to others in affirming ways rather than shaming \cite{ross2019speaking}, however, in excess, this could lead to reducing the opportunity for people to bring more authentic emotions, including negative ones, into conversations \cite{Hohenstein2023-zj}.
Therefore, further research is needed to understand how to best support users in generating authentic yet empathetic and emotionally-aware responses to counter hate speech. 
Moreover, while ease of use is important for counterspeech tools as discussed in Section~\ref{ssec:rq2}, the ease of counterspeaking may lead to more exposures to content that might be harmful, especially if these tools are finding and encouraging users to respond or increasing the speed of hate and counter hate interactions.
To mitigate this issue, design methods to encourage mindfulness such as design frictions that intentionally slows down interaction \cite{cox2016friction,masrani2023mindful} and reflective designs to encourage reflection on intentions \cite{scott2023intentions,sengers2005reflective} should be further studied to find balance between efficiency and mindfulness in context of counterspeech assistance and be integrated to help users be more mindful.

\subsection{Bias in AI Against Marginalized Groups}
Our participants also highlighted biases of AI against minorities as one of the concerns (Section~\ref{sssec:doubts}), which numerous studies have pointed out as a key issue in AI systems \cite{bolukbasi2016man,blodgett-etal-2021-stereotyping,pmlr-v81-buolamwini18a}.
AI systems, especially language models, are often trained on Western-centric data \cite{santy-etal-2023-nlpositionality}, and often exclude and filter out marginalized identities during preprocessing \cite{dodge-etal-2021-documenting}.
These opaque design decisions can thus negatively impact marginalized communities, for example, through disparate performance \cite{sap2019risk,groenwold2020investigating,ZACK2024e12} and operationalized stereotypes \cite{blodgett-etal-2021-stereotyping,cheng-etal-2023-marked}.

As hate speech often targets marginalized communities \cite{UNreport2021}, mitigating AI biases is a key consideration for designing a system to alleviate its impact.
Therefore, methods to address this critical challenge should be further explored when building AI counterspeech tools. 
One strategy could involve amplifying marginalized voices and reflecting more diverse values through different aggregation or careful data collection \cite{sorensen2024roadmap,Gordon_2022,sap2022annotators}.
Furthermore, ongoing efforts to audit biases in AI systems \cite{blodgett-etal-2021-stereotyping,luccioni2023stable,cheng-etal-2023-marked} can surface different biases and enhance their mitigation.
Additionally, it's crucial to delve deeper into the multidimensionality of identities \cite{hanna2020race} and their representation \cite{chasalow2021representativeness} concerning AI systems and downstream tasks.
\newText{
\section{Limitations and Future Work}

\paragraph{Survey Participant Demographics}
Both our survey and interview studies and their analysis are limited to the answers from our participants.
Our survey study was conducted with North American (U.S. and Canada) participants and were overwhelmingly from United States.
Moreover, due to our choice of platform (MTurk), our results might not be representative of populations that are less familiar with technology.
The demographics of our survey participants was also skewed especially in racial and sexual identities as more than 80\% identified as white or Caucasian and heterosexual. 
While our interview participants were from more diverse geographic regions, the interviews were all conducted in English, limiting our results to English-speaking countries. 
Therefore, our results may not generalize to populations outside the ones listed.
These demographic limitations are especially important to be explored by future works given the disproportionate effect of technology on minority groups \cite{scheuerman2019gender,schlesinger2018race} and frequent targeting of minority groups in hate speech \cite{paz2020hate,sahaRiseFearSpeech2023}. 

\paragraph{Definition of AI}
To focus on collecting diverse ideas for AI usages, we did not restrict the definition of AI when asking participants to envision AI tools.
However, this could have lead to different understanding of what constitutes AI, especially based on participants' familiarity with AI \cite{long2020literacy}. 
While collecting levels of experience with AI systems from the participants could have provided more insight into their answers, our work did not cover the interaction between different levels of experience with AI and differences in participant answers.  
This intersection of AI experiences and envisioning of AI tools would be an interesting future work.

\paragraph{Impact of AI and Counterspeech on Minority Groups}
In addition to expanding our study to more diverse demographics, future works should consider studying the specific impact of AI-driven counterspeech systems on marginalized communities. 
AI systems in general are known to exacerbate societal biases \cite{Shams2023} and lead to further marginalization and amplification of existing structural inequalities \cite{sap2019risk,crawford2023}.
This risk is particularly salient in AI counterspeech settings which aim to help and not further victimize hate-targeted communities, and there has already been evidence of backfiring of AI systems (e.g., refusing to talk about race \cite{schlesinger2018race} or misgendering the user \cite{scheuerman2019gender}).
Therefore, continued effort in developing participatory methods to ethically engage different marginalized communities and intersections of such communities \cite{smith2020race,haimson2023transtech,harrington2023intersection} are essential.
}{R3}

\paragraph{Limited Context of Counterspeech}
Future works could also explore counterspeech with a more global lens to understand a wider set of barriers and their interaction with AI assistance. 
\newText{Especially under authoritarian contexts, counterspeech, typically thought of as combating hate, could be used to suppress speech or dissidents \cite{gunitsky_2015}. 
Therefore, understanding the positionality of counterspeakers and their cultural contexts would be an important area of future work and an important ethical consideration toward understanding dual use of AI-driven counterspeech tool.
Additionally, our work focused on text-based counterspeech largely agnostic to choice of community or online platform; future works could explore various modalities, platforms (e.g., TikTok), and communities for counterspeech.}{R3}
We also scoped our current work's focus on countering hate speech. 
However, other forms of online harm, such as fake news, could also be addressed by counterspeech. 
Future works could explore countering fake news and other forms of harm through AI tools. 
Additionally, in this paper, we lay out several design considerations that should be explored by future works.
An iterative design process should take place to implement these considerations to co-design a counterspeech tool to empower and support users. 

\paragraph{Limitation of Methods}
While we used several measures to ensure quality of answers (Appendix~\ref{app:qual}), due to decentralized nature of crowdsourcing-based studies, it is difficult to guarantee that data came from reliable and expected sources.
Further, crowdworkers may use AI-based tools such as ChatGPT to perform annotation~\cite{veselovsky2023artificial}, which can be difficult to distinguish from human responses~\cite{clark-etal-2021-thats}.
Therefore, in human annotation of bot-like responses, we could have allowed both false negatives and false positives, resulting in limitations in the internal validity of the data.

\section{Conclusion}
This work explored the experiences, needs, and concerns of activist counterspeakers and lay participants towards participatory AI for counterspeech. 
Our findings surfaced a theory of counterspeaking process, along with barriers at each step and motivations that drive this process. 
Our work highlighted the tension between the barriers (e.g., limited resources, lack of training, unclear impact, and personal harms) and motivations (e.g., moral duty and positive impact) and several ways that AI tool could help lower the barriers to counterspeech. 
Furthermore, we also surfaced concerns over the use of AI tools for counterspeech in authenticity, agency, and functional doubts. 

Our findings reveal a considerable gap between current direction of research for AI assistance in counterspeech and an empowering assistive tool for users as the negative impact of AI involvement are not fully considered or addressed.
To close this gap, we make several design recommendations connecting our findings to previous works to inform an empowering, user-focused, design of counterspeech AI tools. 
We provide recommendations focusing on transparency to build trust and authenticity in online communication, on design methods to encourage deliberation and moral agency, and on mindful designs to promote mental health.
Our discussion also raises questions about how to best reduce effort and barriers of counterspeech without detracting from meaningful communication and connection.
Thus, our work calls for further exploration and co-design of AI tools for counterspeech that addresses the participants' concerns to empower users in building safer and healthier online spaces. 

\begin{acks}
We would like to thank our anonymous reviewers for their careful consideration and detailed feedback to improve our work. 
We thank the interview participants and MTurk workers for their contribution to counterspeech and our work. 
Additionally, we are thankful to Kathleen Fraser, Isar Nejadgholi, and Svetlana Kiritchenko for a fruitful discussion. 
Jenny T. Liang was supported by the National Science Foundation under grants DGE1745016 and DGE2140739.
\end{acks}

\bibliographystyle{ACM-Reference-Format}
\bibliography{custom}

\appendix
\section{Survey}

\subsection{Response Quality Checks}
\label{app:qual}

\paragraph{Pre-qualification Process}
To ensure the quality of our results, we used a pre-qualification process to prevent fraudulent responses.
The pre-qualification process included three questions:
\begin{enumerate}
\item Humans are mammals.\\
  How true do you think is the above statement?
\item People are right handed.\\
  What percentage of people do you think are right handed?  
\item Penguins can't fly.\\
  What percentage of penguins do you think can't fly?
\end{enumerate}

The workers answer using a slider with percentage ranging 0 to 100 or 11 point likert scale. 
We accept the answer to correct for each question if they answer 1) 10, 2) greater than or equal to $50\%$, and 3) 10. 
We consider a worker qualified if they score 3 on this task. 
The workers were paid $0.22$ USD for the qualification task.

\paragraph{Bot Detection}
We used Google's reCaptcha V2\footnote{\url{https://developers.google.com/recaptcha/docs/display}} and V3\footnote{\url{https://developers.google.com/recaptcha/docs/v3}}. 
Moreover, one of the authors manually annotated answers for bot-like behaviors looking for responses that were repetitive, off-topic, or non-sensical.

\subsection{Participant Demographics}
\label{app:dem}
Participant demographics are shown in Table~\ref{tab:demographics}.
We asked participant's age, race, transgender identity, gender identity, sexuality, religion, political leaning, education, and country of residence. 
Our participants were largely from the U.S. and white with many having bachelor's degree or some college experience.
As we filtered for north American (U.S. and Canada) residents on MTurk, other countries of residence might indicate erroneous reporting. 

\begin{table*}[!ht]
    \centering
    \footnotesize
    \subfloat[Age]{
        \begin{tabular}{l | l}
        Option & Response (\%) \\
        \hline
        18-24 years old         & 02.9240\\
        25-34 years old         & 34.7953\\
        35-44 years old         & 34.5029\\
        45-54 years old         & 14.3275\\
        55-64 years old         & 09.9415\\
        65+ years old           & 02.9240\\
        Prefer not to disclose  & 00.5848\\
        \end{tabular}
    }\hfill
    \subfloat[Race]{
        \begin{tabular}{l | l}
        Option & Response (\%) \\
        \hline
        White or Caucasian                                                   & 82.6979\\
        White or Caucasian,Asian,Native Hawaiian or Other Pacific Islander   & 00.2933\\
        White or Caucasian,Asian                                             & 01.7595\\
        White or Caucasian,Other                                             & 00.5865\\
        White or Caucasian,American Indian/Native American or Alaska Native  & 01.1730\\
        White or Caucasian,Black or African American                         & 00.2933\\
        White or Caucasian,Black or African American,Other                   & 00.2933\\
        White or Caucasian,Native Hawaiian or Other Pacific Islander         & 00.2933\\
        American Indian/Native American or Alaska Native                     & 00.5865\\
        Asian                                                                & 05.8651\\
        Black or African American                                            & 04.6921\\
        Prefer not to say                                                    & 00.8798\\
        Other                                                                & 00.5865\\
        \end{tabular}}\hfill
   \subfloat[Transgender]{
        \begin{tabular}{l | l}
        Option & Response (\%) \\
        \hline
        Yes                     & 02.3460\\
        No                      & 96.4809\\
        Prefer not to disclose  & 01.1730\\
        \end{tabular}}
    \subfloat[Gender]{
        \begin{tabular}{l | l}
        Option & Response (\%) \\
        \hline
        Man                                                   & 56.4327\\
        Woman                                                 & 41.2281\\
        Two-spirit,Woman                                      & 00.2924\\
        Genderqueer or gender fluid                           & 00.2924\\
        Additional gender category/identity   & 00.2924\\
        Prefer not to disclose                                & 01.4620\\
        \end{tabular}}\hfill
    \subfloat[Sexuality]{
        \begin{tabular}{l | l}
        Option & Response (\%) \\
        \hline
        Straight (heterosexual)          & 85.0877\\
        Bisexual                         & 06.4327\\
        Bisexual,Pansexual               & 00.5848\\
        Asexual                          & 02.0468\\
        Asexual,Straight (heterosexual)  & 00.2924\\
        Lesbian                          & 01.1696\\
        Gay                              & 01.1696\\
        Pansexual                        & 01.1696\\
        Questioning or unsure            & 00.2924\\
        Prefer not to disclose           & 01.7544\\
        \end{tabular}}\hfill
    \subfloat[Religion]{
        \begin{tabular}{l | l}
        Option & Response (\%) \\
        \hline
        Atheist                   & 16.9591\\
        Christian                 & 39.1813\\
        Agnostic                  & 16.3743\\
        Catholic                  & 15.4971\\
        Jewish                    & 01.1696\\
        Buddhist                  & 01.1696\\
        Hindu                     & 00.5848\\
        Muslim                    & 00.2924\\
        Nothing in particular     & 04.3860\\
        Prefer not to disclose    & 02.3392\\
        Something else, Specify:  & 02.0468\\
        \end{tabular}}\hfill
    \subfloat[Political Leaning]{
        \begin{tabular}{l | l}
        Option & Response (\%) \\
        \hline
        Strongly liberal        & 22.2222\\
        Liberal                 & 32.4561\\
        Moderate                & 18.7135\\
        Conservative            & 16.6667\\
        Strongly conservative   & 08.4795\\
        Prefer not to disclose  & 01.4620\\
        \end{tabular}}\hfill
    \subfloat[Education]{
        \begin{tabular}{l | l}
        Option & Response (\%) \\
        \hline
        Bachelor’s degree                                                     & 57.6023\\
        Some college, but no degree                                           & 13.1579\\
        Graduate or professional degree  & 07.6023\\
        High school diploma or GED                                            & 10.8187\\
        Associates or technical degree                                        & 09.3567\\
        Some high school or less                                              & 00.8772\\
        Prefer not to say                                                     & 00.5848\\
        \end{tabular}}\hfill
    \subfloat[Country of Residence]{
        \begin{tabular}{l | l}
        Option & Response (\%) \\
        \hline
        United States of America                              & 97.9472\\
        Namibia                                               & 00.2933\\
        Canada                                                & 00.5865\\
        United Kingdom of Great Britain and Northern Ireland  & 00.2933\\
        India                                                 & 00.2933\\
        Albania                                               & 00.2933\\
        Argentina                                             & 00.2933\\
       \end{tabular}}
    \hfill
     \caption{Demographics of survey participants. All the heuristics are reported as percentages.}
     \label{tab:demographics}
\end{table*}

\section{Analysis Results}

The codes developed from participant responses following the methods in Sections~\ref{sssec:interview-analysis-method} and \ref{sssec:survey-analysis-method} are listed in Tables~\ref{tab:cs-barriers-codes}, \ref{tab:ai-tools-codes}, and \ref{tab:ai-concerns-codes}.
The high level themes as discussed in \ref{ssec:theory-of-counterspeaking}, Section~\ref{ssec:barriers} and \ref{ssec:rq3} are indicated using specific icons for visibility.
Counterspeech barriers could be categorized to high level themes of \textit{limited resources} {\color{resources}\faClock}, \textit{lack of training} {\color{training}\faPen}, \textit{unclear impact} {\color{impact}\faHeartBroken}, and \textit{personal harms} {\color{personalharms}\faUserInjured}. 
Moreover, motivations are mapped to intrinsic motivation of \textit{moral duty} \faThumbsUp[regular]\xspace and extrinsic, \textit{positive impact} \faHandHoldingHeart.
Some concerns about AI were related to motivations and are indicated using the same icon. 
However, some additional themes emerged in long-term concerns \faSignOut*\xspace and functional doubts \faQuestionCircle[regular].
\label{app:analysis}
\begin{table*}[!htbp]
    \centering
    \footnotesize
    \begin{tabularx}{\textwidth}{ lXX }
    \toprule
    \textbf{Code} & \textbf{Description}  &
    \textbf{Representative Quote}\\
    \hline 
    \hline
    \rowcolor{headergray}
    \multicolumn{3}{l}{\cellcolor{headergray}{\textbf{\textit{Barriers to Counterspeech}}}} \\
    {\color{resources}\faClock}\xspace\textsf{\myuline{Resources}} & Financial or people resources are limited, especially time it takes to do counterspeech & \textit{``[The biggest challenge is] time. We're doing a max already but for security reasons we cannot be too big. This is not our job.''}\\
    \rowcolor{pastelgray}{\color{resources}\faClock}\xspace\textsf{\myuline{{Finding hate speech}}} & Takes time to find hate speech & \textit{``It's so time consuming looking for articles.''} \\
    {\color{training}\faPen}\xspace\textsf{\myuline{Training}} & People don't have the training & \textit{``People don't always know what to say.''}\\
    \rowcolor{pastelgray}{\color{impact}\faHeartBroken}\xspace\textsf{\myuline{Reach}} & It's hard to reach people, which is discouraging & \textit{``When you feel unheard and it’s like I’m doing this for nothing - it’s not really getting the word out - it’s frustrating''} \\
    {\color{personalharms}\faUserInjured}\xspace\textsf{\myuline{Risk}} & There are risks of online or offline attacks & \textit{``And the risk – the risk is real. I don’t use anonymous posting. In our community, there is fear and so it’s risky when you put yourself out there.''} \\
    \rowcolor{pastelgray}{\color{personalharms}\faUserInjured}\xspace\textsf{\myuline{Mental health}} & It is too hard on mental health (stress or boredom) & \textit{``You are just overwhelmed with what you are seeing.''}\\	
    \hline
    \hline
    \rowcolor{headergray}
    \multicolumn{3}{l}{\cellcolor{headergray}{\textbf{\textit{Counterspeaker Motivations}}}} \\
    {\faThumbsUp[regular]}\xspace\textsf{\myuline{Right}} & It's the right thing to do / civic responsibility & \textit{``I think it’s because it’s the right thing to do – I feel that at least I tried. ''}\\
    \rowcolor{pastelgray}\faHandHoldingHeart\xspace\textsf{\myuline{Impact}} & Seeing evidence of successful impact & \textit{``When you can see the comment section change. When you can see other non-members speaking out. We take screen shots and save our successes.''}\\
    \faHandHoldingHeart\xspace\textsf{\myuline{Scale - small}} & Quotes about the individual-level impact & \textit{``Probably when we get in real time that we've helped someone, helped someone who had maybe been reading the comments and had been upset by them, they say something like, `oh thank goodness, I was in a pit of dispair before seeing your comment.'''}\\
    \bottomrule
\end{tabularx}
    
    \caption{Codes developed from analysis of interview responses discussing counterspeech barriers and motivations.}
    \label{tab:cs-barriers-codes}
\end{table*}
\begin{table*}[htpb]
    \centering
    \footnotesize
    \begin{tabularx}{\textwidth}{ lcXX }
    \toprule
    \textbf{Code} & & \textbf{Description}  &
    \textbf{Representative Quote} \\
    \hline 
    \hline
    \rowcolor{pastelgray}
    \multicolumn{4}{l}{\cellcolor{pastelgray}{\textbf{\textit{Activist Counterspeakers}}}} \\
    \textsf{\myuline{Time}} 
 & {\color{resources}\faClock} & References to time or efficiency  & \textit{``Anything that could help us be more efficient - help us produce more.''} \\
    \rowcolor{pastelgray}\textsf{\myuline{Finding - AI}} & {\color{resources}\faClock} &	AI would help by locating hateful speech & \textit{``I've spent up to two hours looking for good actions, so that would be super helpful. ''} \\
    \textsf{\myuline{Scale - big}} & {\color{impact}\faHeartBroken} & AI would help scale the work of counterspeakers & \textit{``A tool that would amplify voice against hate speech. That would assist in amplifying counterspeech and helping it reach the target audience.''}\\
    \hline
    \hline
    \multicolumn{4}{l}{\cellcolor{pastelgray}{\textbf{\textit{Lay-users}}}} \\
    \textsf{\myuline{Efficiency}} & {\color{resources}\faClock} & The user wants to use the tool to save time. & \textit{``It would be make responding to hatred so much easier if I could just click a few boxes and let an AI do the work. Even though it won't change the hater's heart, it would provide a counter to their hate speach.''} \\
    \rowcolor{pastelgray}\textsf{\myuline{AI-better}} & {\color{resources}\faClock} & The user thinks that AI would be better thant they are. & \textit{``[An AI] could give me a constructive framework for a much more impactful response than I could otherwise generate on my own.''}\\
    \textsf{\myuline{Capability-dependent}} & {\color{resources}\faClock} & The user wants to explore the capabilities of the tool before deciding. & \textit{``I would be willing to see the suggestion that the AI offered and decide whether or not to use it.''}\\
    \rowcolor{pastelgray}\textsf{\myuline{Information}} & {\transparent{0.8}\color{resources}\faClock\llap{\transparent{0.7}\color{training}\faPen}} & The user thinks a tool would be helpful to compile information to counter hate. & \textit{``if it was fact based i sure would use it, since i feel we all can have our own oppinions''}\\
    \textsf{\myuline{Guidance}} & {\color{training}\faPen} & The user would use the tool to get guidance on how to respond effectively: formulating response and creating more diverse response in a more collaborative way, or to help them understand the hate or detection of hate. & \textit{``I would potentially use it because it could give me a constructive framework for a much more impactful response than I could otherwise generate on my own''}\\
    \rowcolor{pastelgray}\textsf{\myuline{Emotions}} & {\color{training}\faPen} & The user wants help with regulating their emotions to communicate clearly or with effectively communicating user's emotions. & \textit{``... It would help me to stay calm and collected. When I am faced with hateful speech, I can sometimes get emotional. This can make it difficult for me to respond effectively. The AI tool would help me to stay calm and collected, so that I could focus on responding to the hateful speech in a thoughtful and reasoned way. It would help me to feel more confident in my responses...''}\\
    \textsf{\myuline{Empowerment}} & {\color{impact}\faHeartBroken} & The user feels empowered by being able to speak up in addressing hate speech to create a positive impact. & \textit{``Because it would help me speak up more.''}\\
    \rowcolor{pastelgray}\textsf{\myuline{Reduce-stress}} & {\color{personalharms}\faUserInjured} & The user feels that having the tool could help reduce stress while responding to hate speech. & \textit{``It might be less stressful to use than making a more personal comment.''}\\
    \textsf{\myuline{AI-proxy}} & {\color{personalharms}\faUserInjured} & The user would rather have the AI get involved, rather themselves (often under the guise of anonymity) either in responding or reporting. & \textit{``To be honest it sort of makes me feel like I have some plausible deniability if an issue arises. In a worst case scenario I would be able to `blame' it on the AI.''}\\
    \hline
    \hline
    \multicolumn{4}{l}{\cellcolor{pastelgray}{\textbf{\textit{Lay-users - AI Tools}}}} \textit{``''}\\ 
    \textsf{\myuline{Existing-technology}} & & The user refers to existing technology a specific AI tool that current exists in the market. & \textit{``I use most of the time ChatGPT.''}\\
    \rowcolor{pastelgray}\textsf{\myuline{Report}} & {\color{resources}\faClock} & The user would like a system that can automatically or with minimal input report hate speech. & \textit{``[An AI] that identifies the hate speech and remove or block the comment.''}\\
    \textsf{\myuline{Response-support}} & {\color{resources}\faClock} & The user wants an AI tool that suggests or automatically replies with a response to hate speech, which could also be refined by the user with minimal input that is well-written and thoughtful. The user usually wants efficient and time saving support with minimal engagement and is easy to use. & \textit{``The one which suggest the reply in very decent manner.''}\\
    \rowcolor{pastelgray}\textsf{\myuline{Factual}} & {\transparent{0.8}\color{resources}\faClock\llap{\transparent{0.7}\color{training}\faPen}} & The user expresses that it would be beneficial to have an assistive technology that can gather factual information to formulate arguments or fact-check hate speech. The users also want help in creating responses that rational, intellectual, and logical arguments. & \textit{``I would use it if it gave out information that was correct and if it was reliable.''}\\
    \textsf{\myuline{Collaborative}} & {\color{training}\faPen} & The user wishes to have more collaborative interaction to improve their responses. Some examples of interactions include correction to their written response such as grammar, emotional, or factual and checking their own biases.  & \textit{``Something I could be "checked" on, making sure MY post wasn't also toxic.''}\\
    \rowcolor{pastelgray}\textsf{\myuline{Effective-communication}} & {\color{training}\faPen} & The user would like to use an AI tool that is sensitive to human emotions while addressing hate speech, and is capable of expressing nuanced emotions. The user wants support communicating clearly with the understanding of emotional, human factors focusing on meaning and impact. & \textit{``An AI assistance that is nice and helps alleviate the situation.''}\\
    \textsf{\myuline{Aligned}} & {\color{impact}\faHeartBroken} & The user would like an AI tool that is personally and/or culturally aligned and provide responses just like how they would or in an unbiased way. & \textit{``One that is trained off my data and personality that I approve of.''}\\
    \rowcolor{pastelgray}\textsf{\myuline{Protective}} & {\color{personalharms}\faUserInjured} & The user would like an AI tool that will protect them from retaliation often through anonimity. & \textit{``I would like an AI tool that could prepare a message...[avoids] making myself a target.''}\\
    \bottomrule
\end{tabularx}
    
    \caption{Codes developed from analysis of responses showing openness to adopt AI tools in SQ14 and in the interview studie as well as responses to SQ15. The icons indicate the theme of barriers relevant to the code. The codes are separated into three sections: benefits identified by activist counterspeakers, benefits identified by lay-users, and AI tools discussed by lay-users. The colors of icons were chosen to match Figure~\ref{fig:theme-interaction}.}
    \label{tab:ai-tools-codes}
\end{table*}
\begin{table*}[htpb]
    \centering
    \footnotesize
    \begin{tabularx}{\textwidth}{ lcXX }
    \toprule
    \textbf{Code} & & \textbf{Description}  &
    \textbf{Representative Quote} \\
    \hline 
    \hline
    \rowcolor{pastelgray}
    \multicolumn{4}{l}{\cellcolor{pastelgray}{\textbf{\textit{Activist CounterSpeakers}}}} \\
    \textsf{\myuline{Authenticity - strategy}} & \faHandHoldingHeart & Worries that inauthentic counterspeech would not be credible & \textit{``The moment we deploy this online, a lot of people who share hateful content and know a lot about tech will recognize it."} \\
    \rowcolor{pastelgray}\textsf{\myuline{Real}} & \faSignOut* & Quotes comparing AI to what is "real" & \textit{``We do not need, neither for us nor for the haters, the possibility to create a fake sentiment and take away our voices. It will boil down to who has the money to pay it more."} \\
    \textsf{\myuline{Long-term}} & \faSignOut* & Concerns that AI counterspeech has long-term negative consequences & \textit{``Really using the bot at all is tricky. You aren't inspiring real people to participate. If we are actually going to make change, we need those people to be engaged. We need people to get involved in their communities."} \\
    \rowcolor{pastelgray}\textsf{\myuline{Counterspeaker}} & \faThumbsUp[regular] & Counterspeakers are aided by doing counterspeech & \textit{``And there is something incredibly magic about turning something really hateful in the other direction. You feel you aren't hopeless or helpless."} \\
    \textsf{\myuline{Becoming the monster}} & \faThumbsUp[regular] & Troll farms are bad. Would we become just as bad by using a counterspeech bot?  & \textit{``I can see the appeal to that for sure, but I think that it takes out the human component. We're kind of no better than what is being used."} \\
    \rowcolor{pastelgray}\textsf{\myuline{Emotional Intelligence}} & \faQuestionCircle[regular] & Emotional intelligence, empathy, you need a human, authenticity  & \textit{``The process itself I find very satisfying. Having a sense of not being alone... For me, what's really touching is that someone out there is just there to support me. I'm not the only one who thinks this."} \\
    \textsf{\myuline{Functionality - technical}} & \faQuestionCircle[regular] & Skepticism that AI counterspeech would work  & \textit{``I’m not sure about it getting the facts right. It’s not good for fact checking."} \\
    \hline
    \hline
    \rowcolor{pastelgray}
    \multicolumn{4}{l}{\cellcolor{pastelgray}{\textbf{\textit{Lay-users}}}} \\
    \textsf{\myuline{Authenticity}} & \faHandHoldingHeart & The user expresses concern that what AI communicates is not their own words and would not represent what they are thinking especially their intentions (alignment) or would be considered inauthentic focusing on "who" is behind counterspeech. & \textit{``Using AI is too impersonal and it sounds very generic."} \\
    \rowcolor{pastelgray}\textsf{\myuline{Engaging-not-helpful}} & \faSignOut* & The user believes engaging with people who espouse hate speech is not helpful in reducing that behavior or do not want hate speech getting more attention. These users sometimes express that they would engage if they knew that it would make an impact. & \textit{``I don't think it matters if I get help with what I want to say if it's just falling on deaf ears."} \\
    \textsf{\myuline{Avoidance}} & \faSignOut* & The user rather wishes to avoid hate speech rather than engage and availability of an AI tool will not change this. & \textit{``I don't engage in any online hate/drama. I just scroll right through."} \\
    \rowcolor{pastelgray}\textsf{\myuline{Agency}} & \faThumbsUp[regular] & The user does not want AI help especially because they are able to perform the task themselves. The user prefers humans to respond focusing on "how" counterspeech is generated. & \textit{``I believe in stating things that I feel not what an AI tells me to feel."} \\
    \textsf{\myuline{Capability-doubts}} & \faQuestionCircle[regular] & The user expresses that they do not think that AI response will have an impact or it will contain other functionality problems.  & \textit{``AI tools are often wrong and I wouldn't want it's bias' to affect what I am posting."} \\
    \rowcolor{pastelgray}\textsf{\myuline{Become-target}} & \faQuestionCircle[regular] & The user does not want to become the target of hate. & \textit{``Having an AI help me write a response would not keep people from sending me hateful replies. I cannot handle that."} \\
    \bottomrule
    \end{tabularx}
    \caption{Codes from analysis of responses resistant to adopting AI tools from SQ14 and concerns raised by interview participants. The icons indicate relevant themes of motivations that are negatively affected and new themes of functionality doubts and long-term impact. The codes are presented in two sections: responses from activist counterspeakers and from lay-users.}
    \label{tab:ai-concerns-codes}
\end{table*}

\end{document}